\documentclass[iop]{emulateapj}

\usepackage{natbib}
\usepackage{url}
\usepackage{graphicx}
\usepackage{amssymb}
\usepackage{amsmath}

\def\modif#1{{#1}}

\shorttitle{Forming chondrules in impact splashes}
\shortauthors{Dullemond et al.}

\begin{document}

\title{Forming chondrules in impact splashes\\ II Volatile retention}

\author{Cornelis Petrus Dullemond, Daniel Harsono, Sebastian Markus Stammler}
\affil{Institute for Theoretical Astrophysics, Heidelberg University,
Albert-Ueberle-Strasse 2, 69120 Heidelberg, Germany}

\author{Anders Johansen}
\affil{Lund Observatory, Department of Astronomy and Theoretical
Physics, Lund University, Box 43, 22100 Lund, Sweden}

\begin{abstract}
  Solving the mystery of the origin of chondrules is one of the most elusive
  goals in the field of meteoritics. Recently the idea of planet(esimal)
  collisions releasing splashes of lava droplets, long considered out of
  favor, has been reconsidered as a possible origin of chondrules by several
  papers. One of the main problems with this idea is the lack of
  quantitative and simple models that can be used to test this scenario by
  directly comparing to the many known observables of chondrules. In Paper I
  of this series we presented a simple thermal evolution model of a
  spherically symmetric expanding cloud of molten lava droplets that is
  assumed to emerge from a collision between two planetesimals. The
  production of lava could be either because the two planetesimals were
  already in a largely molten (or almost molten) state due to heating by
  $^{26}$Al (e.g.~Sanders et al.~2005, 2012), or due to impact jetting at
  higher impact velocities (Johnson et al.~2015). In the present paper,
  number II of this series, we use this model to calculate whether or not
  volatile elements such as Na and K will remain abundant in these droplets
  or whether they will get depleted due to evaporation. The high density of
  the droplet cloud (e.g.\ small distance between adjacent droplets) causes
  the vapor to quickly reach saturation pressure and thus shutting down
  further evaporation. We show to which extent, and under which conditions,
  this keeps the abundances of these elements high, as is seen in
  chondrules. We find that for most parameters of our model (cloud mass,
  expansion velocity, initial temperature) the volatile elements Mg, Si and
  Fe remain entirely in the chondrules. \modif{The Na and K abundances
    inside the droplets will initially stay mostly at their initial values
    due to the saturation of the vapor pressure, but at some point start to
    drop due to the cloud expansion. However, as soon as the temperature
    starts to decrease, most or all of the vapor recondenses again. At the
    end the Na and K elements retain most of their initial abundances,
    albeit occasionally somewhat reduced, depending on the parameters of the
    expanding cloud model. These findings appear to be qualitatively
    consistent with the analysis of Semarkona Type II chondrules by Hewins,
    Zanda \& Bendersky (2012) who found evidence for sodium evaporation
    followed by recondensation.}
\end{abstract}


\section{Introduction}
\label{sec-introduction}
Chondritic meteorites consist for a large part of 0.1 to 1 millimeter size
silicate spherules that were once molten lava droplets. These
so-called chondrules are much larger than the dust grains in the
interstellar medium, and thus must have formed in the protoplanetary disk
out of which our solar system was created. So far there is no conclusive
evidence as to what was the energy source that produced these 2000 K hot
lava droplets. But from the analysis of the textures of these chondrules it
is generally concluded that the cooling process was fast: a matter of hours
(Hewins et al.\ 2005; see also references in Morris \& Desch 2010). There
are abundant theories of what could be the origin of chondrules, among them
are nebular shocks (e.g.~Hood \& Hor\'anyi, 1991; Desch \& Connolly 2002;
Ciesla \& Hood 2002; see also Stammler \& Dullemond 2014), nebular lightning
(Horanyi et al.~1995; Eisenhour \& Buseck 1995; Gibbard et al.~1997); the
X-wind model (Shu et al.~2001) and flash heating by energy dissipation in
current sheets forming in MHD turbulence (Hubbard et al.~2012). There are
energy conservation analyses showing that the required energy is a
non-negligible fraction of the accretion energy of the protoplanetary disk
(King \& Pringle 2010). Finally, Jacquet et al.~(2012) include mixing
processes within the disk in their analysis to see how chondrules produced
in different locations disperse and form chondrites.

The alternative theory that chondrules might have originated as a result of
planetesimal collisions producing a splash of molten droplets that cooled
down to become chondrules (e.g.~Urey 1953; Kieffer 1975; Zook 1980) has long
been dismissed. However, recently interest in this idea has been revived
(Sanders \& Taylor 2005; Hevey \& Sanders 2006; Asphaug et al.~2011; Sanders
\& Scott 2012; Fedkin et al.~2012; Fedkin \& Grossman 2013; Johnson et
al.~2015). One of the main problems with the planetesimal collision scenario
is that it takes large impact velocities ($\gtrsim 3\cdots 5$ km/s) to
generate enough impact heat to melt the rock out of which the planetesimals
are made. During the gas-rich phases of the protoplanetary disk (lasting a
few million years) one would expect that due to friction with the gas the
planetesimals have only small eccentricities and inclinations, and thus
would collide at much slower speeds, meaning that no melt is produced. Zook
(1980) argued that this problem can be solved if one takes into account that
short-lived radionuclides (in particular $^{26}$Al) can heat up
planetesimals of $\gtrsim 10$ km radius beyond the solidus temperature,
essentially turning them into spheres of magma with a crust (Hevey \&
Sanders 2006; Sanders \& Scott 2012). A collision would release the magma
into a spray of lava droplets, even if the collision speed is
moderate. Asphaug et al.~(2011) argued that one can calculate the droplet
radius to be consistent with observed sizes if the planetesimals are at
least 10 km or more in size.

A potential problem with the pre-molten planetesimals is that they might
differentiate if the melting is sufficient, the gravity strong enough and
convective mixing not efficient enough. This would mean that the melt they
would release would be very non-solar. Recently Johnson et al.~(2015)
therefore revisited the original high-impact-speed idea in which the cloud
of droplets is caused by ``impact jetting''. In this model, initially
studied in this context by Kieffer (1975), pre-melting is not necessary
since the impact velocity is high enough to create the melt though
shock-heating at the interface between the colliding bodies. The produced
melt will then, under high pressure, escape as a sheet of ``jets'' to the
side. Johnson et al.~show that for realistic collision velocities of about 3
km/s only a small fraction of the mass of the impacting bodies ends up in a
jet, essentially implying that chondrules are merely a by-product of
colliding planets. They calculate, using Monte Carlo calculations of
planetesimal populations including the effect of eccentricity damping by the
nebular gas, that nevertheless a sufficient number of collisions would occur
to account for the total mass in chondrules in the asteroid belt.

Either way (pre-melting or high-velocities), the impact origin of chondrules
is being re-investigated. A problem with testing the impact hypothesis is
that quantitative models of the splash are scarce, because the process is so
complex (see e.g.\ Asphaug et al.~2011 and Johnson et al.~2015 for
hydrodynamic simulations), making it very hard to make quantiative model
predictions for quantities such as cooling times and volatile element
retention, which are the kind of data that are obtained from meteoritic
studies.

In a previous paper we tried to remedy this by presenting a very simple
model of a spherically symmetric ballistically expanding cloud of hot lava
droplets that radiatively cool through time-dependent radiative transfer
(Dullemond, Stammler \& Johansen 2014, henceforth Paper I). This thermal
model provides droplet number densities and temperatures as a function of time
and location within the cloud. In the present paper we will use this model
to make model predictions for the abundances of volatile elements in
chondrules, and compare this to generic properties of these abundances in
chondrites.

The reason for focusing on volatile elements is because the abundances of
volatile elements such as Na and K in most chondrules is fairly close to
solar (see e.g.~Fedkin \& Grossman 2013 for a discussion). This is
surprising given the high temperatures needed to melt rock to form lava. A
millimeter-size lava droplet in vacuum at a temperature of ca.~2000 K would
evaporate and lose its Na and K within less than a minute. The fact that we
do not see strong depletion of these elements in most chondrules suggests
that the chondrules were at those high temperatures only extremely briefly
(tens of seconds at most), suggesting rapid heating and subsequent rapid
cooling of the chondrules. This appears to be inconsistent with constraints
on cooling rates, which suggest cooling times of the order of
hours. Moreover, if evaporation was efficient, one would expect strong
isotopic fractionation to occur, which is not observed in chondrules (Cuzzi
\& Alexander 2006).

A very natural way to prevent the loss of volatile elements is, however, if
the chondrules are very close together during their high-temperature
formation phase: each chondrule will then have only a limited volume of
``private'' vacuum around it in which it can outgas its volatiles, and the
vapor will then quickly reach saturation pressure. Or in other words: a
chondrule will absorb what another nearby chondrule will evaporate. This
idea was proposed by Cuzzi \& Alexander (2006) and Alexander et al.~(2008)
to explain the lack of volatile element depletion and fractionation in
chondrules. They conclude that chondules formed a very dense cloud of
hundreds of kilometers across before the chondrules cooled down and became
solid. Such high-density clouds of solids are not unnatural in
protoplanetary disks. For instance, the streaming instability has been shown
to produce extremely high concentrations of solids (Johansen \&
Youdin~2007).  It is, however, very hard to imagine any nebular scenario
(processes happening in the gas-and-dust protoplanetary disk) which could
heat up a large and very dense cloud of proto-chondrules to temperatures of
$\sim$ 2000 K in a matter of minutes to hours. The density of such a cloud
would be much larger than the gas density of the protoplanetary disk, so
that the gas of the disk would not likely be able to release enough energy
in a short enough time to heat the proto-chondrules from nebular
temperatures (few hundred Kelvin) to melting tempeature (ca 2000 K). 

Fedkin \& Grossman (2012), Hewins, Zanda \& Bendersky (2012) and Fedkin et
al.~(2013) also analyzed the vapor saturation scenario and concluded that
impact splashes, in which droplets of molten lava are ejected from an impact
event, are the most natural scenario to create these high-density,
high-temperature, short-lived environments.

It is the purpose of this series of papers to test this idea with a simple
model of a ballistically expanding cloud of droplets that cools through
emission of infrared radiation (Paper I). Here (Paper II) we add an
evaporation/condensation model to it, but we deliberately keep this
evaporation/condensation model simple. The geo/cosmochemical problem of
evaporation/condensation of minerals is actually quite complex, and the
above mentioned papers treat this in a much more detailed way. Our
simplified treatment aims merely at investigating whether the general
scenario is feasible, given a ballistically expanding cloud of droplets.

We find that as the cloud of hot lava droplets moves away from the impact
site and expands, the vapor quickly reaches saturation pressure. If the
cloud is large enough then the vapor diffusion time scale would be larger
than the expansion time so that during the expansion process the vapor is
not escaping. The total amount of volatile elements in the vapor phase is
initially much less than the amount that remains in the solid phase, but
under some conditions shortly before the radiative cooling sets in, a
non-negligible fraction of the volatile elements are in the gas phase. As
the cloud then cools, the vapor will recondense onto the chondrules. Before
the expanding cloud becomes too tenuous for recondensation to be effective
all the volatiles have recondensed.

The structure of this paper is as follows. For details on the thermal model
we refer to Paper I, but we will give a very brief summary in Section
\ref{sec-cloud-model}. Then we will discuss the time-dependent evaporation
and condensation model in Section \ref{sec-evapcond}, which is a simplified
version of a more detailed model described in the appendices
\ref{sec-equil-pvap-metal-oxides}, \ref{sec-peq-mgfenak}
\ref{sec-time-dep-evapcond-metal-oxides} and \ref{app-janafberman}. In this
section we also describe the results. Finally in Section
\ref{sec-discussion} we will discuss the results and come to a conclusion.

\section{Expanding cloud model}
\label{sec-cloud-model}
The goal of the simple model of Paper I was to compute the density and
temperature of the cloud of lava droplets as a function of time after the
planetesimal collision. The dynamics was assumed to be extremely simple: a
spherically symmetric homogeneous cloud expanding with a constant rate while
it is moving away from the impact site. The radius of the cloud is thus a
linear function of time $t$ since the collision:
\begin{equation}
R_{\mathrm{cloud}}(t) = v_{\mathrm{exp}}t
\end{equation}
where $v_{\mathrm{exp}}$ is the expansion velocity of the cloud. The 
density of chondrules in the cloud is then
\begin{equation}\label{eq-rho-afo-time}
\rho_{\mathrm{cloud}}(t) =\frac{3M_{\mathrm{cloud}}}{4\pi R_{\mathrm{cloud}}^3(t)} 
\end{equation}
where $M_{\mathrm{cloud}}$ is the total mass of all chondrules in the cloud.
The temperature of the lava droplets declines with time as soon as they can
radiatively cool. However, right after the onset of expansion the cloud is
still very optically thick, meaning that the radiation cannot escape the
cloud fast enough. The temperature thus initially stays constant. Only after
some time $t_{\mathrm{cool}}$ the optical depth drops below some critical
value (which is still well above unity) below which the radiative diffusive
energy loss kicks in and the temperature starts to drop. In Paper I we
calculated this time-dependent radiative cooling numerically with a
time-dependent radiative transfer algorithm. We found that these numerical
results can be understood and fitted with a simple analytic model. We
briefly describe the model here, and refer to Paper I for details. We
consider chondrules (lava droplets) of radius $a_{\mathrm{chon}}$ and
material density $\xi_{\mathrm{chon}}$. We take $\xi_{\mathrm{chon}}=3.3$
g/cm$^3$ for our model. The mass of a chondrule is then $m_{\mathrm{chon}} =
\tfrac{4\pi}{3}\xi_{\mathrm{chon}} a_{\mathrm{chon}}^3$. The number density
of chondrules is $n_{\mathrm{chon}}(t) =
\rho_{\mathrm{cloud}}(t)/m_{\mathrm{chon}}$. With the simplifying assumption
of zero albedo the absorption opacity is then $\kappa =
\tfrac{3}{4}(\xi_{\mathrm{chon}}a_{\mathrm{chon}})^{-1}$ (geometric
opacity). The optical depth from the center of the cloud to the edge is
$\tau(t) = \rho_{\mathrm{cloud}}(t) \kappa R_{\mathrm{cloud}}(t)$.

In this paper we will present our results based on a set of fiducial models
as well as parameter scans. The parameters of the fiducial models are listed
in Table \ref{tab-model-param}. Note that instead of using
$M_{\mathrm{cloud}}$ as the parameter determining the amount of mass in the
cloud of droplets, we instead use $R_{\mathrm{melt,0}}$: the equivalent
radius of a sphere of melt with mass $M_{\mathrm{cloud}}$, defined through
\begin{equation}\label{eq-def-rmelt0}
M_{\mathrm{cloud}}=\frac{4\pi}{3}\xi_{\mathrm{chon}}R_{\mathrm{melt,0}}^3
\end{equation}
Using $R_{\mathrm{melt,0}}$ as parameter is more intuitive than
$M_{\mathrm{cloud}}$, which is the only reason for using
$R_{\mathrm{melt,0}}$ instead of $M_{\mathrm{cloud}}$.

\begin{table}
\begin{center}
\begin{tabular}{|cccc|c|}
\hline
\hline
Model & $R_{\mathrm{melt,0}}$ & $v_{\mathrm{exp}}$ & $T_0$ & $t_{\mathrm{cool}}$ \\
\hline
F1         & 1 km            & 100 m/s  & 2000 K   & 26 min\\
F2         & 0.1 km          & 1000 m/s  & 2000 K  & 16 sec\\
F3         & 10 km           & 100 m/s  & 2000 K   & 7 h\\
F4         & 0.01 km         & 1000 m/s  & 2000 K  & 1 s\\
\hline
\hline
\end{tabular}
\end{center}
\caption{\label{tab-model-param}The model parameters for the main
  (``fiducial'') models presented in this paper. The cloud mass $M_{\mathrm{cloud}}$
  can be derived from $M_{\mathrm{cloud}}=\tfrac{4\pi}{3}\xi_{\mathrm{chon}}R_{\mathrm{melt,0}}^3$. 
}
\end{table}

\begin{figure}
\begin{center}
\includegraphics[width=0.49\textwidth]{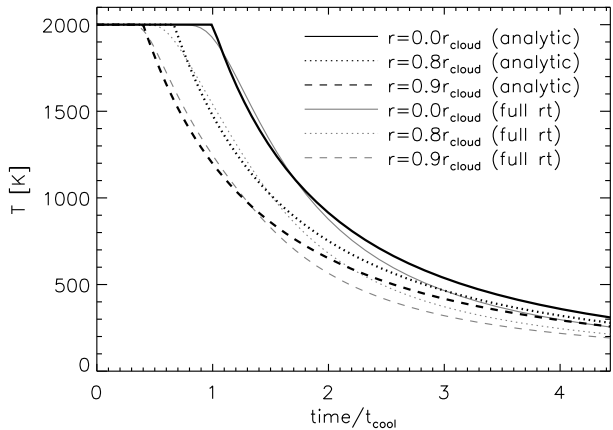}
\end{center}
\caption{\label{fig-temp-afo-time-analytic}\modif{The temperature evolution
    of the expanding cloud model for the fiducial model F1 (see Table
    \ref{tab-model-param}). Shown in black are the estimates according to
    the} analytical estimate of
  Eqs.~(\ref{eq-t-analytic},\ref{eq-t-analytic-zero}). In grey are, for
  comparison, the full radiative transfer solutions from Paper I. The
  temperatures are shown at three positions in the cloud: the center, at
  80\% of the radius and at 90\% of the radius (near the surface of the
  cloud). The time scale is scaled to $t_{\mathrm{cool}}$
  (Eq.~\ref{eq-tcool}). The analytic solutions are only valid as long as
  \modif{the optical depth} $\tau_{\mathrm{cool}}\gg 1$.}
\end{figure}

In Paper I we found that the temperature of the chondrules at the center of
the cloud remains at the initial temperature $T_0$ for a duration
$t_{\mathrm{cool}}$ given by:
\begin{equation}\label{eq-tcool}
t_{\mathrm{cool}}=\left(\frac{1}{5}\frac{3}{(4\pi)^2}
\frac{M_{\mathrm{cloud}}^2c_m\kappa}{v_{\mathrm{exp}}^4\sigma
T_0^3}\right)^{1/5}
\end{equation}
where $c_m\simeq 10^7$ erg g$^{-1}$ K$^{-1}$ is the specific heat capacity
of the chondrule material. For $t>t_{\mathrm{cool}}$ the temperature
drops with time. In Paper I it was found that a reasonable analytic 
description of this temperature decline is:
\begin{equation}\label{eq-t-analytic}
T(t>t_{\mathrm{cool}}) = T_0\left[\frac{3}{5}\frac{t}{t_{\mathrm{cool}}}+
\frac{2}{5}\right]^{-5/3}
\end{equation}
We assume that for $t<t_{\mathrm{cool}}$ we have:
\begin{equation}\label{eq-t-analytic-zero}
T(t<t_{\mathrm{cool}}) = T_0
\end{equation}
The solution Eqs.~(\ref{eq-t-analytic},\ref{eq-t-analytic-zero}) holds true
for the central temperature. But most of the mass of a homogeneous sphere
resides in the outer parts. For the temperature at $r=0.8R_{\mathrm{cloud}}$
and $r=0.9R_{\mathrm{cloud}}$ (roughly the radii of half mass and of 75\%
mass respectively) the best fitting solution to the full radiative transfer
solutions of Paper I are similar to Eq.~(\ref{eq-t-analytic}) but with the
term $2/5$ replaced by $3/5$ and $3.8/5$ respectively. These analytic
estimates of the temperature as a function of time are plotted in
Fig.~\ref{fig-temp-afo-time-analytic}, where also the full radiative
transfer model results are overplotted. \modif{Note that the resulting
  temperature curves at the center and near the surface of the cloud are
  qualitatively the same - the only strong difference being the time of
  onset of the cooling. This means that almost independent of the location
  within the cloud the chondrule undergo a similarly rapid or slow cooling.}

\section{Time-dependent evaporation/condensation}
\label{sec-evapcond}
Now that we know fairly accurately how the temperature of the cloud declines
with time for the given parameters of the model, we will compute the
evaporation and condensation process to find out if the chondrules will lose
their volatile elements, and if yes, by roughly which amount. The main
volatile elements of interest here are Na and K. But we will include other
elements such as Fe, Si and Mg mainly as a background reference, since these
make up the bulk of the mass of the lava droplet.

It should be said that various earlier papes have considered the geochemical
problem of evaporation and condensation of Na and K in an impact scenario in
much more detail and with much more sophistication than we do here, in
particular the paper by Fedkin \& Grossman (2013). The main new aspect we
focus on here is how this works for the analytic cloud expansion model
(Paper I).

Let us consider a volatile material that is initially locked up in the
chondrule and might evaporate out of the chondrule when it is hot and
liquid. We will focus on Na$_2$O, K$_2$O, FeO, SiO$_2$ and MgO.

In Appendices \ref{sec-equil-pvap-metal-oxides}, \ref{sec-peq-mgfenak}
\ref{sec-time-dep-evapcond-metal-oxides} and \ref{app-janafberman} we
describe the evaporation process in detail. We will, however, employ a
simplified model of this process for the time-dependent
evaporation/condensation modeling which we will describe in Section
\ref{sec-simplified-model-evapcond}. The reason for not employing the full
machinery of the evaporation/condensation modeling procedure is that it will
qualitatively not change the results much, but it will make the results much
less clear and reproducable. Our results are therefore meant to give a rough
picture of the process of time-dependent evaporation and condensation of the
volatiles in the cloud of lava droplets. This is, at least for now,
warranted, because the model is anyway highly simplified, both in the
geometry of the cloud as well as in the degree to which laboratory
measurements have so far pinned down the detailed time dependent
evaporation/condensation behavior under all circumstances. The goal of the
present paper is to see how the volatiles behave in the expanding
radiatively cooling cloud of droplets. It is {\em not} the goal to present
an evaporation/condensation model per se. For a much more detailed
evaporation/condensation model we refer to Fedkin \& Grossman (2013).

\subsection{A simplified model of time-dependent evaporation/condensation
of volatile elements from of lava droplets}
\label{sec-simplified-model-evapcond}
Our simplified model is based on the framework described in the appendices
\ref{sec-equil-pvap-metal-oxides}, \ref{sec-peq-mgfenak}
\ref{sec-time-dep-evapcond-metal-oxides} and \ref{app-janafberman}. For
nomenclature: whenever we talk about ``vapor'' we mean the volatile element
in the gas phase, whereas when we talk about ``volatile'' we talk about the
metal oxide itself (be it in the gas phase or inside the liquid chondrule).
We will focus on the volatiles Na$_2$O$(l)$, K$_2$O$(l)$, FeO$(l)$,
SiO$_2$$(l)$ and MgO$(l)$, where $(l)$ stands for the liquid phase. Although
these liquid volatiles are predominantly present in the form of
Na$_2$SiO$_3(l)$, KAlSiO$_4(l)$, Fe$_2$SiO$_4(l)$, SiO$_2(l)$ and
Mg$_2$SiO$_4(l)$, we measure the mass weighted abundance $f_i$ of volatile
$i$ in terms of the evaporating part of this metal oxide, e.g.\ for the
evaporation of Na$_2$SiO$_3$ we measure the mass fraction of Na$_2$O,
because the rest is, upon evaporation, left behind as SiO$_2$. We will
therefore, from here onward, only refer to Na$_2$O etc. We assume that the
droplet is completely liquid, so that the volatile is always perfectly mixed
within the droplet, and no concentration gradients form toward the surface.

In the gas phase $(g)$ the volatiles are present as Na$(g)$, K$(g)$,
Fe$(g)$, SiO$(g)$, Mg$(g)$ and of course the oxygen O$_2$$(g)$ which escapes
along with the sodium, potassium, iron, silicon oxide and magnesium. We
assume that no other gas is present apart from the vapor itself, i.e.\ we
assume that the nebular gas does not have the chance to enter the cloud
before it has cooled down below the solidus.  In that case the oxygen
pressure follows directly from the partial pressures of the volatile
elements (see Eq.~\ref{eq-oxygen-adding-and-sharing}), so we will focus on
Na, K, Fe, SiO and Mg only. The more refractory elements such as Al and Ca
are not included in the model because they will not evaporate under the
conditions treated in this paper.

From the mass weighted abundance $f_i$ of volatile $i$ (where $i$ is the
index of the volatile species: $i=$Na$_2$O, K$_2$O, FeO, SiO$_2$ or MgO) we
can define the total volatile mass inside the drop:
\begin{equation}\label{eq-mass-volatile}
m_{\mathrm{vol},i}=f_i\; m_{\mathrm{chon}}
\end{equation}
where $m_{\mathrm{chon}}$ is the lava droplet (future chondrule) mass.
Since we will see that the most abundant volatiles (SiO$_2$, FeO and MgO)
evaporate only very little, we can safely assume that the radius
$a_{\mathrm{chon}}$ will stay approximately constant throughout the
simulation.

The rate of evaporation/condensation of a volatile per unit surface area of
the lava droplet is given by the Hertz-Knudsen equation
(Eq.~\ref{eq-hertz-knudsen-eq-general}), which in our simplified form reads
\begin{equation}\label{eq-evaprate-hertzknud}
J_i = \frac{\alpha_i}{\sqrt{2\pi m_{\mathrm{vap},i}kT}}
  \big(p^{\mathrm{eq}}_i(f_i,T)-p_i\big)
\end{equation}
where $m_{\mathrm{vap},i}$ is the vapor particle mass, $k$ is the Boltzmann
constant, $\alpha_i$ is the evaporation/condensation coefficient (the
sticking probability of a vapor particle hitting the surface), $p_i$ is the
vapor partial pressure, and $p^{\mathrm{eq}}_i$ is the equilibrium vapor
pressure of vapor species $i$. The partial pressure of vapor species $i$ is
related to the number density $n_i$ of vapor particles via the ideal gas law
$p_i=n_ikT$. We assume that the chondrule is completely liquid, so that the
volatile elements are always perfectly mixed within the droplet, and no
concentration gradients form toward the surface. 

In writing down Eq.~(\ref{eq-evaprate-hertzknud}) we made several
simplifying assumptions. For one, we assumed that the evaporation and
condensation coefficients are equal:
$\alpha_{\mathrm{evap},i}=\alpha_{\mathrm{cond},i}\equiv \alpha_i$, and
constant under all conditions. We take the values inferred by Fedkin et
al.~(2006) from their careful fitting of the Hertz-Knudsen equation to the
experiments. Their values are approximately $\alpha_{\mathrm{Na}}=0.26$,
$\alpha_{\mathrm{K}}=0.13$, $\alpha_{\mathrm{Fe}}=0.23$,
$\alpha_{\mathrm{SiO}}\simeq 0.14$ and $\alpha_{\mathrm{Mg}}\simeq
0.26$. Secondly, the computation of the equilibrium vapor pressures
$p^{\mathrm{eq}}_i$ for all the species is done using the following
simplified procedure:
\begin{enumerate}
\item Before the start of the simulation we compute the functional form of
  the equilibrium pressures for the coupled congruent
  evaporation/condensation problem as a function of temperature, for the
  given initial composition of the lava droplets 
  \modif{(see appendix \ref{sec-peq-mgfenak})}. \modif{We assume here
  that a negligible amount of nebular gas is present around the lava
  droplets, which is a reasonable assumption given that the expanding
  cloud will be very dense and thus initially ``snowplow'' the nebular
  gas. With this assumption the gas between the lava droplets consists 100\%
  of vapor from the lava droplets themselves.} For each species \modif{we represent} this
  function $p^{\mathrm{eq,0}}_i(T)$ by a 3rd order polynomial
  fit:
  \begin{equation}\label{eq-peq0-polyfit}
    \log^{10}\left(\frac{p^{\mathrm{eq,0}}_i(T)}{10^6\,\mathrm{dyne/cm}^2}\right) = 
    q_{i,0} + q_{i,1}\,T + q_{i,2}\,T^2 + q_{i,3}\,T^3
  \end{equation}
  where 1 bar is $10^6$ dyne/cm$^2$. The coefficients $q_{i,k}$ for the
  ``composition 3'' of Yu et al.~(2003) are given in Table
  \ref{tab-eqpress-yu-3}. During the entire simulation these coefficients
  $q_{i,k}$ are kept fixed, but when the temperature $T$ varies, the values
  of $p^{\mathrm{eq,0}}_i(T)$ change according to Eq.~(\ref{eq-peq0-polyfit}).
\item As the mass fractions $f_i$ of the volatiles vary with time, we assume
  that their equilibrium pressures vary linearly \modif{with} $f_i$:
  \begin{equation}\label{eq-peq-linear-with-f}
    p^{\mathrm{eq}}_i(f_i,T) =\frac{f_i}{f_{0,i}} p^{\mathrm{eq,0}}_i(T)
    \equiv f_i\,p^{\mathrm{eq,00}}_i(T)
  \end{equation}
  Here $f_{0,i}$ is the initial value of $f_i$ (i.e.~the initial composition
  of the droplet), and
  $p^{\mathrm{eq,00}}_i(T)$ is defined as $p^{\mathrm{eq,00}}_i(T)\equiv
  p^{\mathrm{eq,0}}_i(T)/f_{0,i}$ and the function $p^{\mathrm{eq,0}}_i(T)$ is
  given by \modif{Eq.~(\ref{eq-peq0-polyfit}).}
  \modif{The interpretation of $p^{\mathrm{eq,00}}_i(T)$ is the equilibrium
  vapor pressure if the surface would consist 100\% of volatile species $i$.
  The linear behavior of Eq.~(\ref{eq-peq-linear-with-f})} 
  appears to be counter to the expected behavior for Na and K,
  where $p^{\mathrm{eq}}_i$ is expected to be proportional to $\sqrt{f_i}$
  (see appendix \ref{sec-peq-mgfenak}). However, in the vacuum evaporation
  experiments of Yu et al.~(2003) an exponential decay of $f_i$ with time
  appears to be observed for Na and K, which suggests a linear behavoir. 
  \modif{See the discussion by Alexander (2001). For simplicity we will
  therefore stick to a} linear dependence of $p^{\mathrm{eq}}_i$ on $f_i$
  \modif{as given in Eq.~(\ref{eq-peq-linear-with-f}). This equation} 
  also implies that
  variations in $f_i$ do not influence $p^{\mathrm{eq}}_{k\neq i}$. Also
  this is not strictly true, but since our results show that $f_{i}$ will
  never become $f_i\ll f_{0,i}$ this simplification is justified as well.
\end{enumerate}

The loss rate of volatile $i$ from the chondrule is
\begin{equation}\label{eq-dmchondt}
\frac{d(m_{\mathrm{vol},i})}{dt} \equiv \frac{d(f_i m_{\mathrm{chon}})}{dt}
\equiv  m_{\mathrm{chon}}\frac{df_i}{dt} = -4\pi a_{\mathrm{chon}}^2J_i m_{\mathrm{vap},i}
\end{equation}
where we assume that the \modif{droplet} radius $a_{\mathrm{chon}}$ \modif{and
mass $m_{\mathrm{chon}}$ do} not change \modif{appreciably with time}.  The
time scale of evaporation into vacuum (i.e. assuming that \modif{the ambient vapor pressure} $p_i=0$) is then
\begin{equation}\label{eq-tevap}
t_{\mathrm{evap},i} \equiv \frac{f_i\,m_{\mathrm{chon}}}{|d(f_im_{\mathrm{chon}})/dt|} 
=\frac{m_{\mathrm{chon}}kT}{4\pi a_{\mathrm{chon}}^2 m_{\mathrm{vap},i}
\alpha_i v_i p^{\mathrm{eq,00}}_i}
\end{equation}
where $f_i$ has dropped out of the equation because of the assumed linear
scaling between the equilibrium vapor pressure and the mass fraction of the
volatile. This means that \modif{when the lava droplet resides} in vacuum 
\modif{its volatile} abundance $f_i(t)$ goes as
\begin{equation}\label{eq-f-afo-t-vacuum}
f_i(t) = f_{0,i}\,e^{-t/t_{\mathrm{evap},i}}
\end{equation}
where $f_{0,i}$ is the abundance at $t=0$. 

However, in our expanding chondrule cloud model we must include saturation
because the chondrules are very close to each other. We assume that the
space between the chondrules is vacuum except for the vapor that just
evaporated from the chondrules. Each chondrule has a volume $V$ around it
which it can fill up with vapor. The idea here is that if a cloud of
chondrules of mass $M_{\mathrm{cloud}}$ has a radius of
$R_{\mathrm{cloud}}$, then the volume per chondrule is:
\begin{equation}\label{eq-volume-per-chondrule}
V = \frac{4\pi}{3}\frac{m_{\mathrm{chon}}\,R_{\mathrm{cloud}}^3}{M_{\mathrm{cloud}}}
\end{equation}
which is then the vacuum space that the vapor from each chondrule can fill
(where we made the assumption that $V\gg (4\pi/3)a_{\mathrm{chon}}^3$, which
is always guaranteed in the parameter range of interest). This assumes, of
course, that the vapor will not hydrodynamically flow out of the cloud.

To get a feeling for the numbers it is instructive to use
Eq.~(\ref{eq-volume-per-chondrule}) to define the typical distance $d$
between neighboring chondrules as twice the radius belonging to the volume
$V$ (Wigner-Seitz radius):
\begin{equation}
d(t)=2\left(\frac{m_{\mathrm{chon}}}{M_{\mathrm{cloud}}}\right)^{1/3}R_{\mathrm{cloud}}(t)
\end{equation}
This distance increases linearly in time. The neighboring chondrules thus
move away from each other at a speed $v_{\mathrm{chonsep}}$ given by
\begin{equation}
v_{\mathrm{chonsep}}=\frac{dd(t)}{dt} = 2\left(\frac{m_{\mathrm{chon}}}{M_{\mathrm{cloud}}}\right)^{1/3}v_{\mathrm{exp}}
\end{equation}
This $v_{\mathrm{chonsep}}$ typically has values of {\em millimeters per
  minute}, i.e.\ these are very low velocities.

At the moment of impact (the start of the expansion of the cloud) we start
with a volatile abundances $f_{0,i}$ inside the chondrule. If we know that
some time later a volatile abundance $f_i$ is left inside the chondrule then
the number density of gas phase vapor particles must be
\begin{equation}\label{eq-nvap}
n_i = \frac{(f_{0,i}-f_i)m_{\mathrm{chon}}}{m_{\mathrm{vap},i}V}
\end{equation}
or equivalently the mass density of vapor particles:
\begin{equation}\label{eq-rhovap}
\rho_i = \frac{(f_{0,i}-f_i)m_{\mathrm{chon}}}{V}
\end{equation}
which is independent of the vapor particle mass $m_{\mathrm{vap},i}$.

In equilibrium (i.e.\ assuming we have infinite time to evaporate and
saturate) the vapor number density $n_i$ of species $i$ must be
\begin{equation}\label{eq-nvap-eq}
n_i = \frac{p^{\mathrm{eq}}_i(f_i)}{kT}
=\frac{f_i\,p^{\mathrm{eq,00}}_i}{kT}
\end{equation}
By combining this equation with Eq.~(\ref{eq-nvap}) we can solve for $f_i$ and
obtain the equilibrium value for the remaining abundance of the volatile
element inside the liquid chondrule:
\begin{equation}\label{eq-fequil}
f_{\mathrm{eq},i} = \left(1+\frac{m_{\mathrm{vap}}Vp^{\mathrm{eq,00}}_i}{m_{\mathrm{chon}}kT}\right)^{-1}f_{0,i}
\end{equation}
This is the saturation value of the abundance of the volatile, given the
volume $V$ around each chondrule. We can also estimate the volume needed
to have 1\% or 50\% of the volatile in the gas phase. We set
$f_{\mathrm{eq},i}=f_{0,i}/100$ or $f_{\mathrm{eq},i}=f_{0,i}/2$ respectively and find
\begin{equation}\label{eq-volfifty}
V_{1\%} = \frac{m_{\mathrm{chon}}kT}{100m_{\mathrm{vap},i}p^{\mathrm{eq,00}}_i}
,\qquad
V_{50\%} = \frac{m_{\mathrm{chon}}kT}{m_{\mathrm{vap},i}p^{\mathrm{eq,00}}_i}.
\end{equation}
We can find the times $t_{1\%}$ and $t_{50\%}$ when this volume is reached
by inserting Eq.~(\ref{eq-volume-per-chondrule}) into
Eq.~(\ref{eq-volfifty}) and solving for $t$:
\begin{eqnarray}
t_{1\%} &=& \left(\frac{M_{\mathrm{cloud}}kT}{100v_{\mathrm{exp}}^3m_{\mathrm{vap},i}p^{\mathrm{eq,00}}_i}\right)^{1/3}
,\nonumber\\
\label{eq-time-50}
t_{50\%} &=& \left(\frac{M_{\mathrm{cloud}}kT}{v_{\mathrm{exp}}^3m_{\mathrm{vap},i}p^{\mathrm{eq,00}}_i}\right)^{1/3}
\end{eqnarray}

The mass loss/gain from/to the chondrule is given by
Eq.~(\ref{eq-dmchondt}).  We can now rewrite Eq.~(\ref{eq-dmchondt}) with
\modif{the equation for the vapor number density} Eq.~(\ref{eq-nvap}), and using \modif{the Hertz-Knudsen equation} (Eq.~\ref{eq-evaprate-hertzknud}), into a
time-evolution equation for the \modif{volatile abundance} $f_i$:
\begin{equation}
\frac{df_i}{dt}=4\pi a_{\mathrm{chon}}^2\frac{m_{\mathrm{vap},i}
\alpha_iv_i}{m_{\mathrm{chon}}}
\left(\frac{(f_{0,i}-f_i)m_{\mathrm{chon}}}{m_{\mathrm{vap},i}V}-\frac{p_i^{\mathrm{eq,00}}f_i}{kT}\right)
\end{equation}
\modif{Using Eq.~(\ref{eq-fequil})} this can be rewritten as
\begin{equation}\label{eq-ode-of-fidt}
\frac{df_i}{dt}=4\pi a_{\mathrm{chon}}^2\frac{\alpha_iv_i}{V}
\left(1+\frac{m_{\mathrm{vap},i}Vp^{\mathrm{eq,00}}_i}{m_{\mathrm{chon}}kT}\right)
(f_{\mathrm{eq},i}-f_i)
\end{equation}
Assuming that between some time $t_1$ and $t>t_1$ the values of $V$, $T$ and
$a_{\mathrm{chon}}$ do not change, the solution is:
\begin{equation}\label{eq-f-afo-t-volume}
f_i(t) = f_{\mathrm{eq},i} + (f_i(t_1)-f_{\mathrm{eq},i})e^{-(t-t_1)/t_{\mathrm{eq},i}}
\end{equation}
with the equilibration timescale $t_{\mathrm{eq},i}$ given by
\begin{equation}\label{eq-timescale-eqpress}
t_{\mathrm{eq},i} = \left[4\pi a_{\mathrm{chon}}^2\frac{\alpha_iv_i}{V}
\left(1+\frac{m_{\mathrm{vap},i}Vp^{\mathrm{eq,00}}_i}{m_ikT}\right)\right]^{-1}
\end{equation}
\modif{As a test, note} that for $V\rightarrow\infty$ \modif{(evaporation into vacuum, no 
saturation)} we obtain
\begin{equation}
\lim_{V\rightarrow\infty} t_{\mathrm{eq},i} \rightarrow t_{\mathrm{evap},i}
\end{equation}
where \modif{the evaporation time scale} 
$t_{\mathrm{evap},i}$ is given by Eq.~(\ref{eq-tevap}). \modif{For this 
extreme case we get $\lim_{V\rightarrow\infty}f_{\mathrm{eq},i}\rightarrow 0$ 
(from Eq.~\ref{eq-fequil}), and
so} solution
Eq.~(\ref{eq-f-afo-t-volume}) then indeed reduces to \modif{the time dependent 
formula for evaporation into vacuum}
Eq.~(\ref{eq-f-afo-t-vacuum}), as expected.  For non-infinite $V$ we have
$t_{\mathrm{eq},i} < t_{\mathrm{evap},i}$.

In the expanding cloud model of Paper I the temperature changes with
time. An analytical approximation to the temperature as a function of time
is given, and also the number density of chondrule droplets (or equivalently
the volume $V$ around each chondrule) as a function of time is a simple
formula. Given that the analytic solution of \modif{the volatile abundance} $f_i(t)$ of
Eq.~(\ref{eq-f-afo-t-volume}) only strictly holds for constant $T$ and
constant $V$, we cannot simply apply Eq.~(\ref{eq-f-afo-t-volume}) to the
expanding cloud model. However, we can apply Eq.~(\ref{eq-f-afo-t-volume})
piecewise for each time step, where $f_{0,i}$ is the abundance at the start
of the time step (time $t_0$) and $f_i(t=t_0+\Delta t)$ is then obtained
from Eq.~(\ref{eq-f-afo-t-volume}). This yields a very stable and easy
method of numerical integration of Eq.~(\ref{eq-ode-of-fidt}). During every
time step we consider $T$ and $V$ constant so that the analytic solution
Eq.~(\ref{eq-f-afo-t-volume}) applies, but for each time step the values of
$T$ and $V$ will be different.  The full procedure is that at each time step
we first calculate the new temperature structure $T(r,t)$, either using the
full radiative transfer method of Paper I or by using its analytic
approximation.  Then we assume that for the entire time step this
temperature and the volume $V$ are constant so that we can use
Eq.~(\ref{eq-f-afo-t-volume}) to compute the new abundances. Then we repeat
this procedure for the next time step etc.

\subsection{Results of the time-dependent evaporation/condensation}
\label{sec-vol-results}
As initial composition we take ``composition 3'' from Yu et al.~(2003).
These values are listed in Table \ref{tab-comp-yu-3}.

\begin{figure*}
\begin{center}
\includegraphics[width=0.48\textwidth]{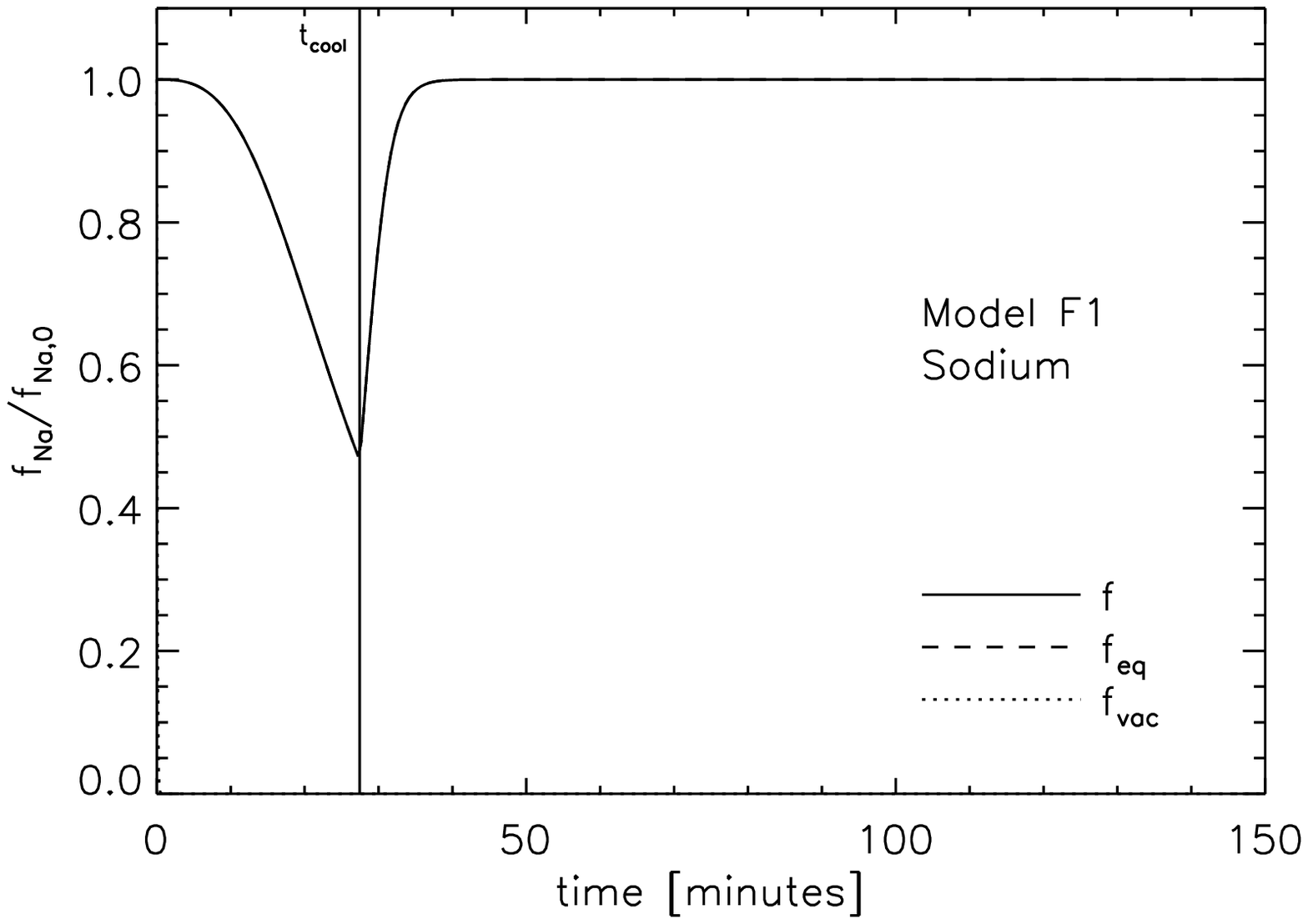}
\includegraphics[width=0.48\textwidth]{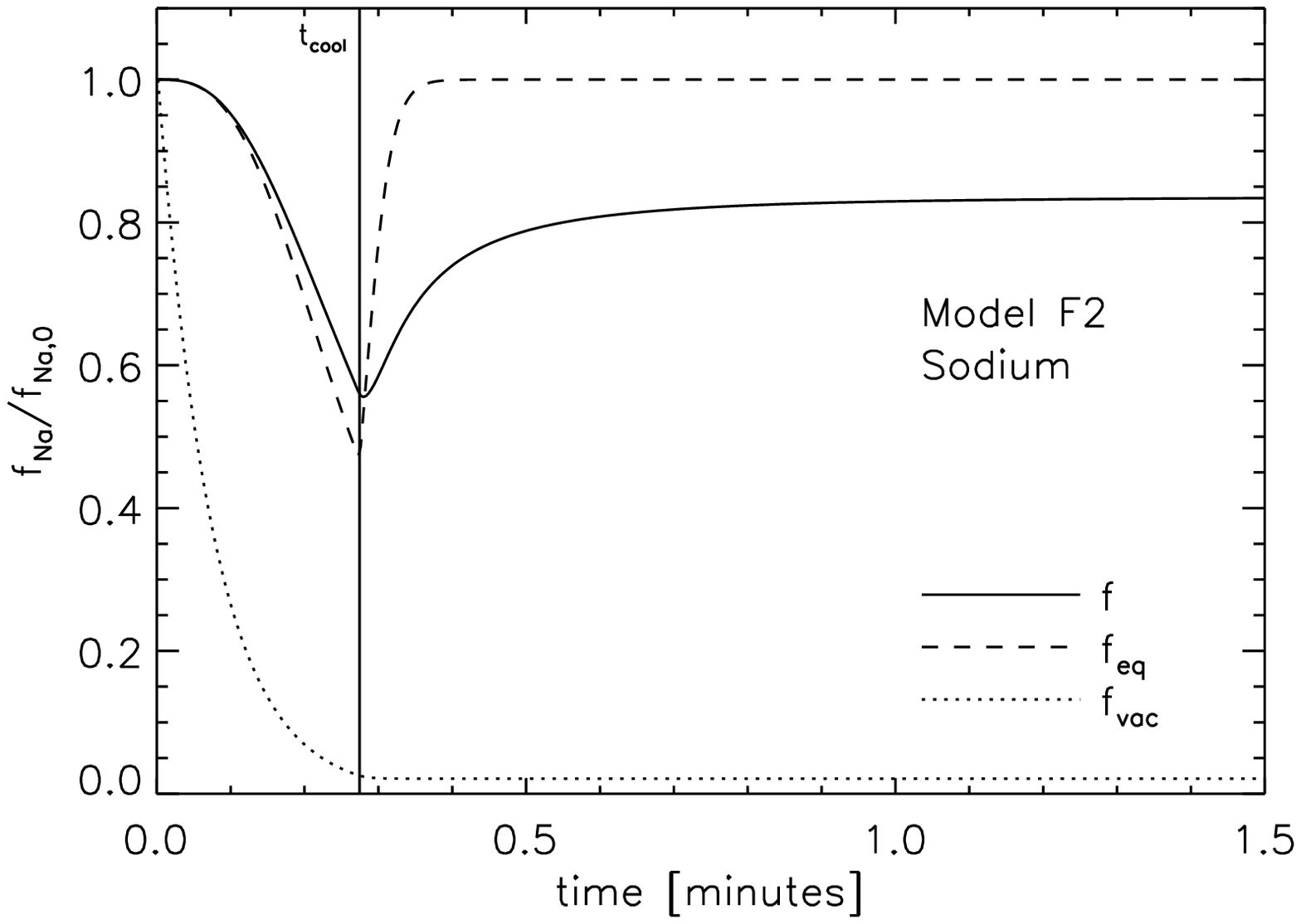}
\end{center}
\caption{\label{fig-abun-feq-fvac}Abundance of Na (relative to the initial
  abundance) as a function of time for models \modif{F1 ($R_{\mathrm{melt,0}}=1\,\mathrm{km}$, 
  $v_{\mathrm{exp}}=0.1\mathrm{km/s}$ - left) and F2 ($R_{\mathrm{melt,0}}=0.1\,\mathrm{km}$, 
  $v_{\mathrm{exp}}=1\mathrm{km/s}$ - right)}. Here
  the temperature model is the analytic model of
  Eqs.~(\ref{eq-t-analytic},\ref{eq-t-analytic-zero}). The solid line is the
  result of the full evaporation/condensation model. The dashed line is the
  equilibrium abundance for the given temperature $T(t)$ and
  volume-per-chondrule $V(t)$ at that time. The dotted line is the abundance
  if only evaporation is included (no condensation, hence no saturation).
  The dashed line can only be distinguished in model F2 because in model F1
  the dashed and solid lines overlap as a result of the equilibration time
  $t_{\mathrm{eq}}$ (Eq.~\ref{eq-timescale-eqpress}) being very small
  compared to $t$ in model F1. The dotted line is also hard to find in model
  F1 because it drops to zero very early on.}
\end{figure*}

\begin{figure*}
\begin{center}
\includegraphics[width=0.48\textwidth]{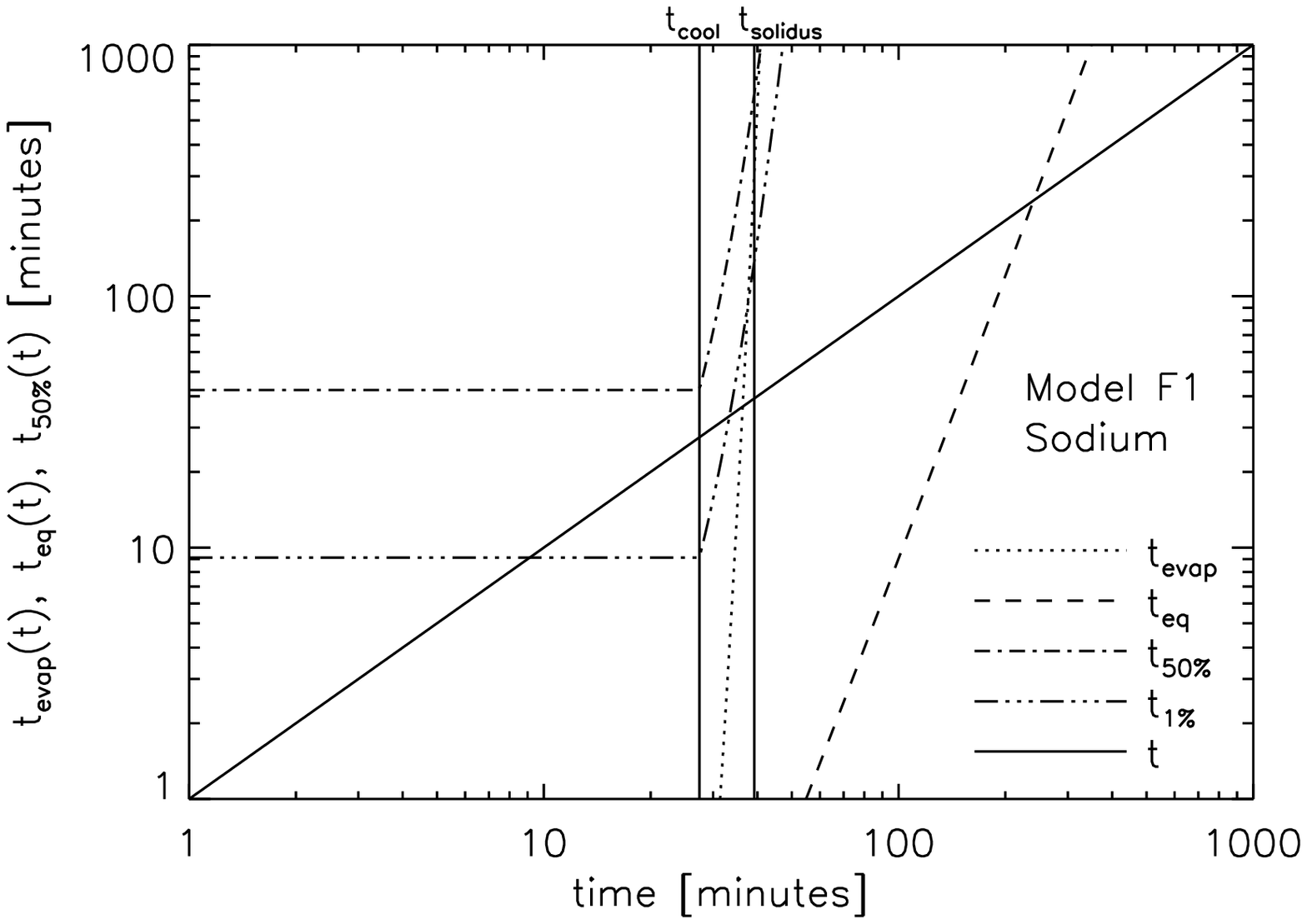}
\includegraphics[width=0.48\textwidth]{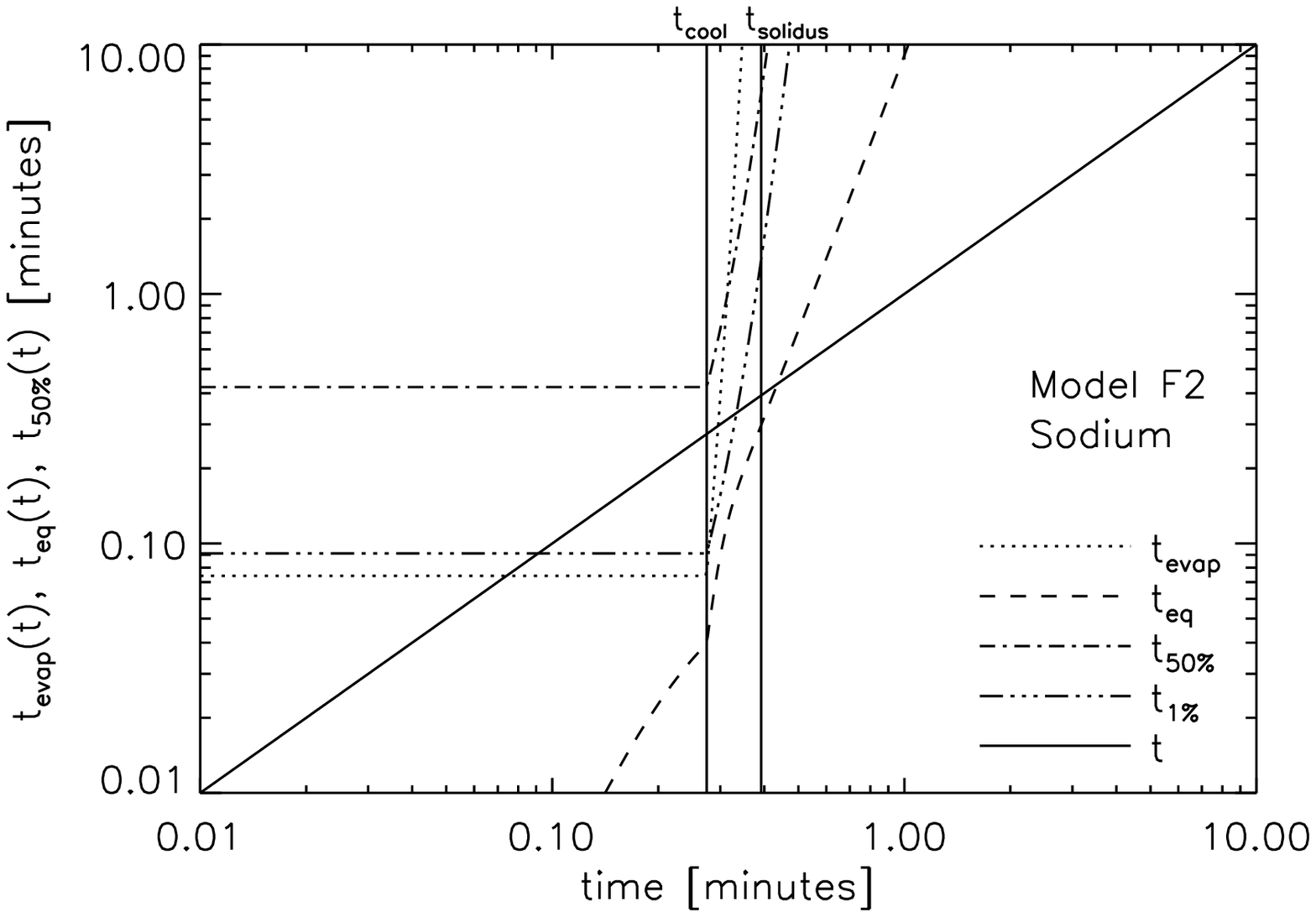}
\end{center}
\caption{\label{fig-evapcond-timescales}The most important time scales as a
  function of time for models \modif{F1 ($R_{\mathrm{melt,0}}=1\,\mathrm{km}$, 
  $v_{\mathrm{exp}}=0.1\mathrm{km/s}$ - left) and F2 ($R_{\mathrm{melt,0}}=0.1\,\mathrm{km}$, 
  $v_{\mathrm{exp}}=1\mathrm{km/s}$ - right)}: the evaporation time
  scale $t_{\mathrm{evap}}$ (dotted line), the equilibration time scale
  $t_{\mathrm{eq}}$, the 50\% evaporation time scale $t_{\mathrm{50\%}}$ and
  the the 1\% evaporation time scale $t_{\mathrm{1\%}}$.  Here the
  temperature model is the analytic model of
  Eqs.~(\ref{eq-t-analytic},\ref{eq-t-analytic-zero}). The time when the
  temperature drop starts ($t_{\mathrm{cool}}$) and the time when the
  solidus temperature of 1400 K is reached ($t_{\mathrm{solidus}}$) are
  marked with vertical lines in the figure. Note that in this figure both
  axes are logarithmic, in contrast to the other figures.}
\end{figure*}

\begin{figure*}
\begin{center}
\includegraphics[width=0.48\textwidth]{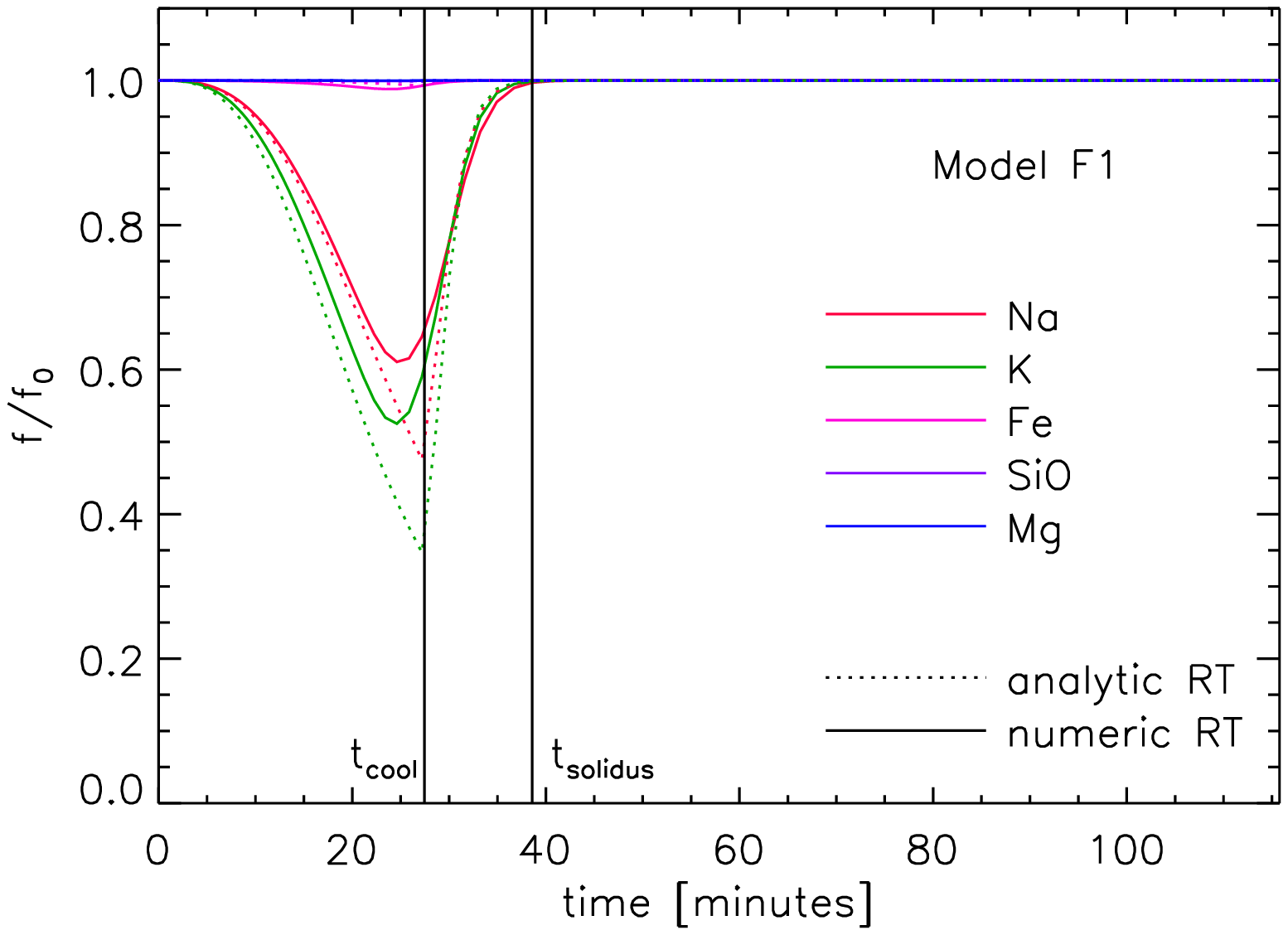}
\includegraphics[width=0.48\textwidth]{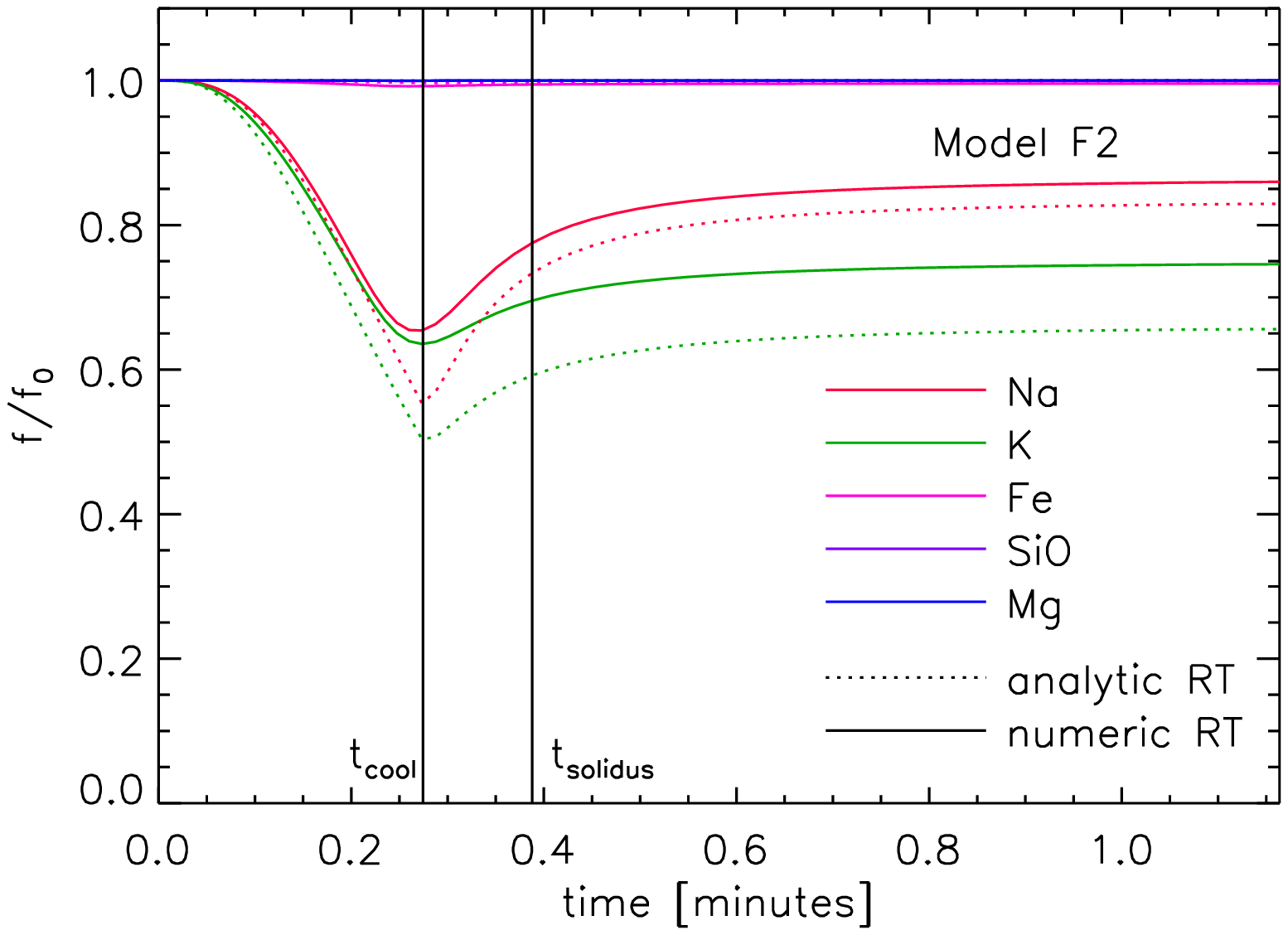}
\end{center}
\begin{center}
\includegraphics[width=0.48\textwidth]{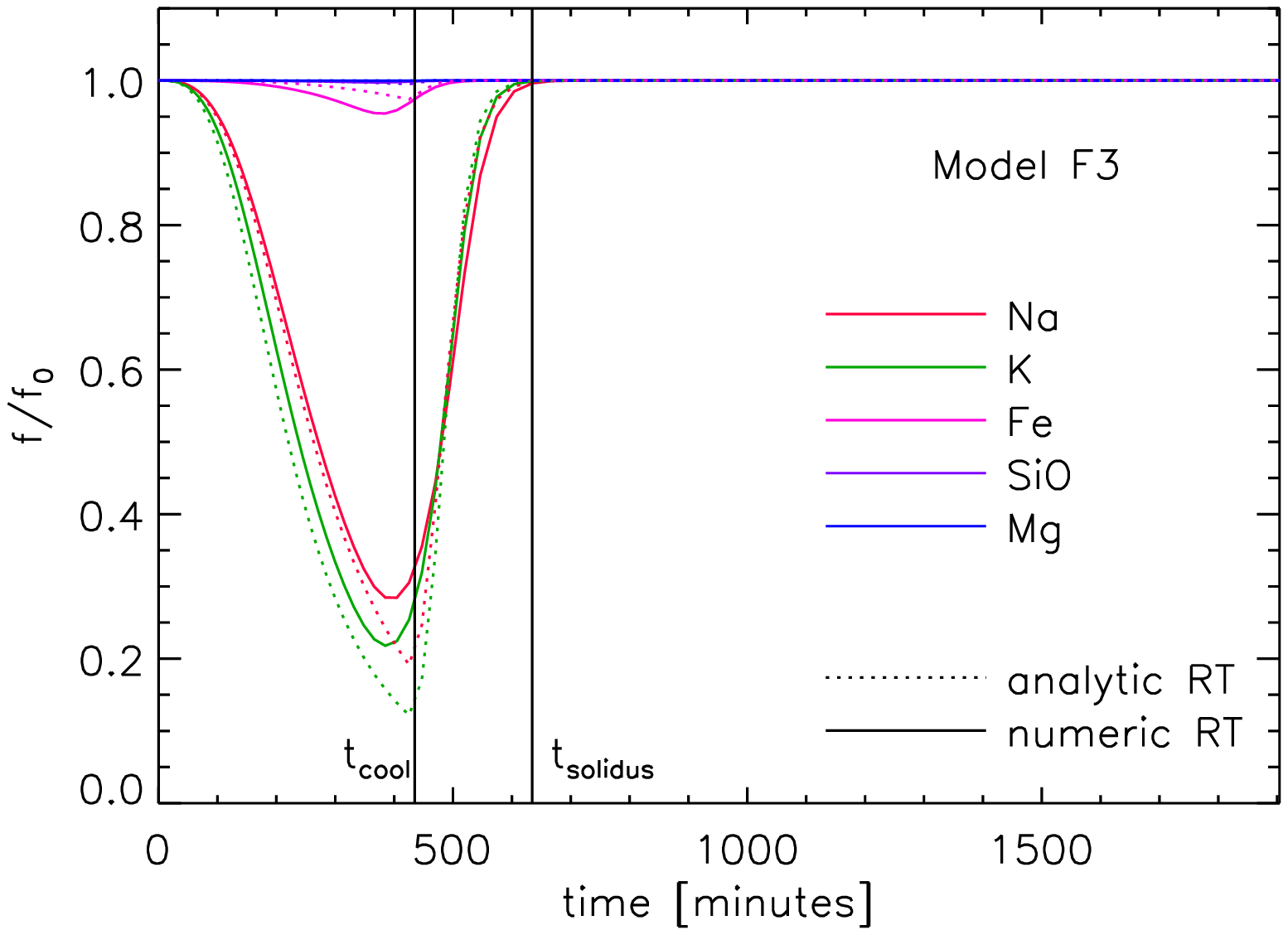}
\includegraphics[width=0.48\textwidth]{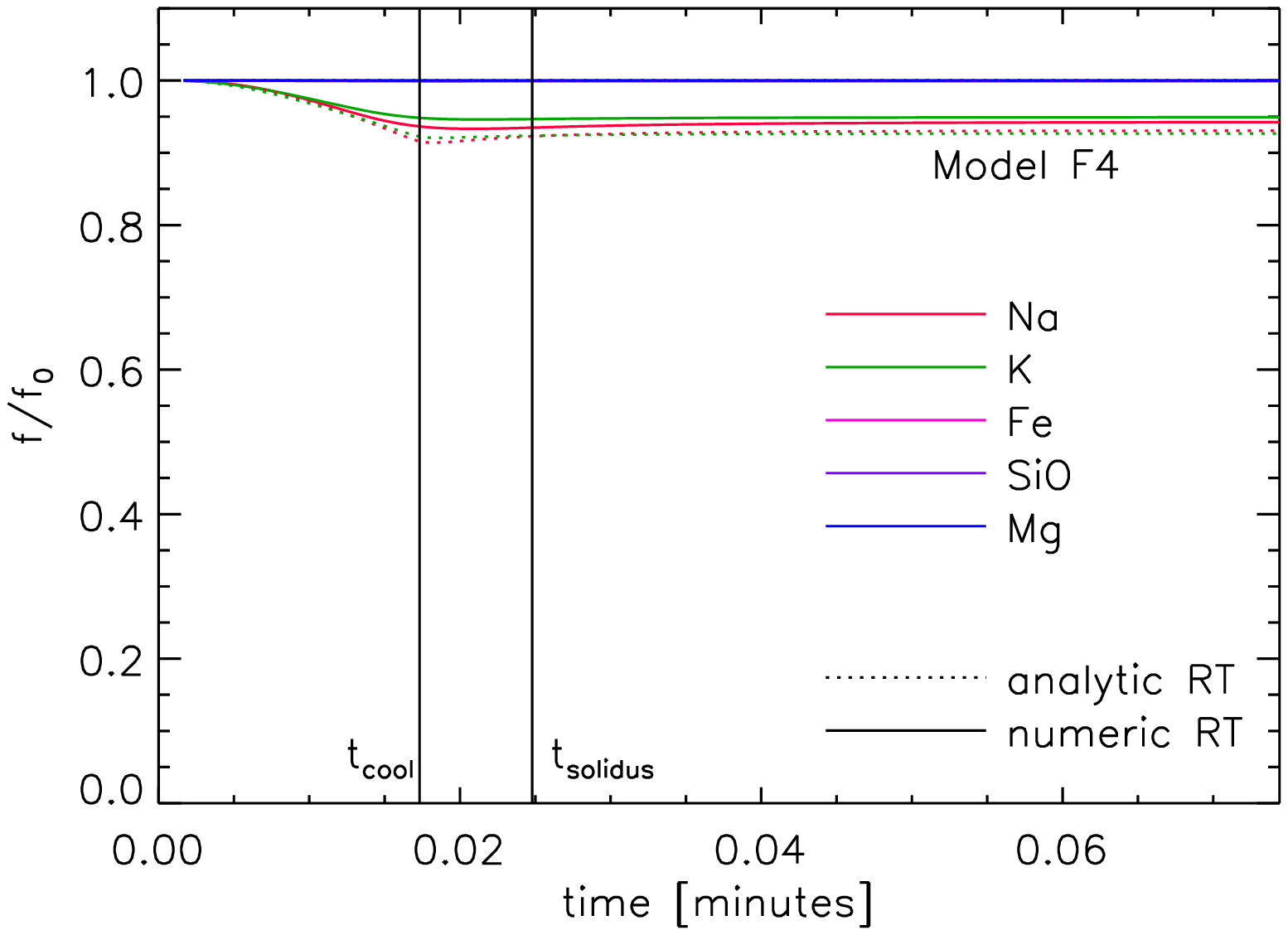}
\end{center}
\caption{\label{fig-abun-an-num}Results of the model 
  \modif{F1 ($R_{\mathrm{melt,0}}=1\,\mathrm{km}$, 
  $v_{\mathrm{exp}}=0.1\mathrm{km/s}$), F2 ($R_{\mathrm{melt,0}}=0.1\,\mathrm{km}$, 
  $v_{\mathrm{exp}}=1\mathrm{km/s}$), F3 ($R_{\mathrm{melt,0}}=10\,\mathrm{km}$, 
  $v_{\mathrm{exp}}=0.1\mathrm{km/s}$) and F4 ($R_{\mathrm{melt,0}}=0.01\,\mathrm{km}$, 
  $v_{\mathrm{exp}}=1\mathrm{km/s}$)}
  (left-to-right, top-to-bottom): the abundances of the volatiles as a
  function of time at the center ($r=0$) of the cloud. Overplotted are the
  models when using the analytic temperature solution (dotted) and the full
  radiative transfer temperature solution (solid).}
\end{figure*}

\begin{figure}
\begin{center}
\includegraphics[width=0.48\textwidth]{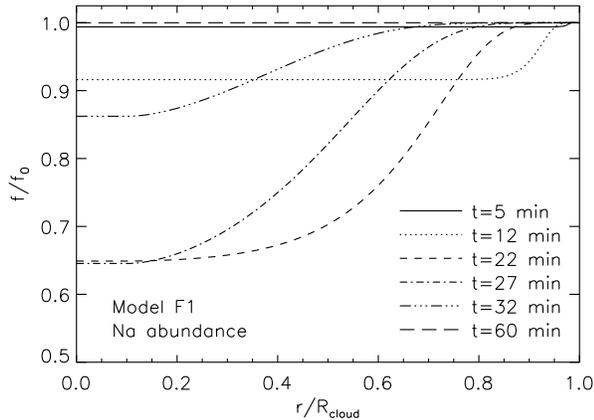}
\end{center}
\caption{\label{fig-abun-na-afo-r}Result of model F1 \modif{($R_{\mathrm{melt,0}}=1\,\mathrm{km}$, 
  $v_{\mathrm{exp}}=0.1\mathrm{km/s}$)} for Na, plotted as a
  function of $\eta=r/R_{\mathrm{cloud}}$ for several time instances.}
\end{figure}

\begin{figure*}
\begin{center}
\includegraphics[width=0.48\textwidth]{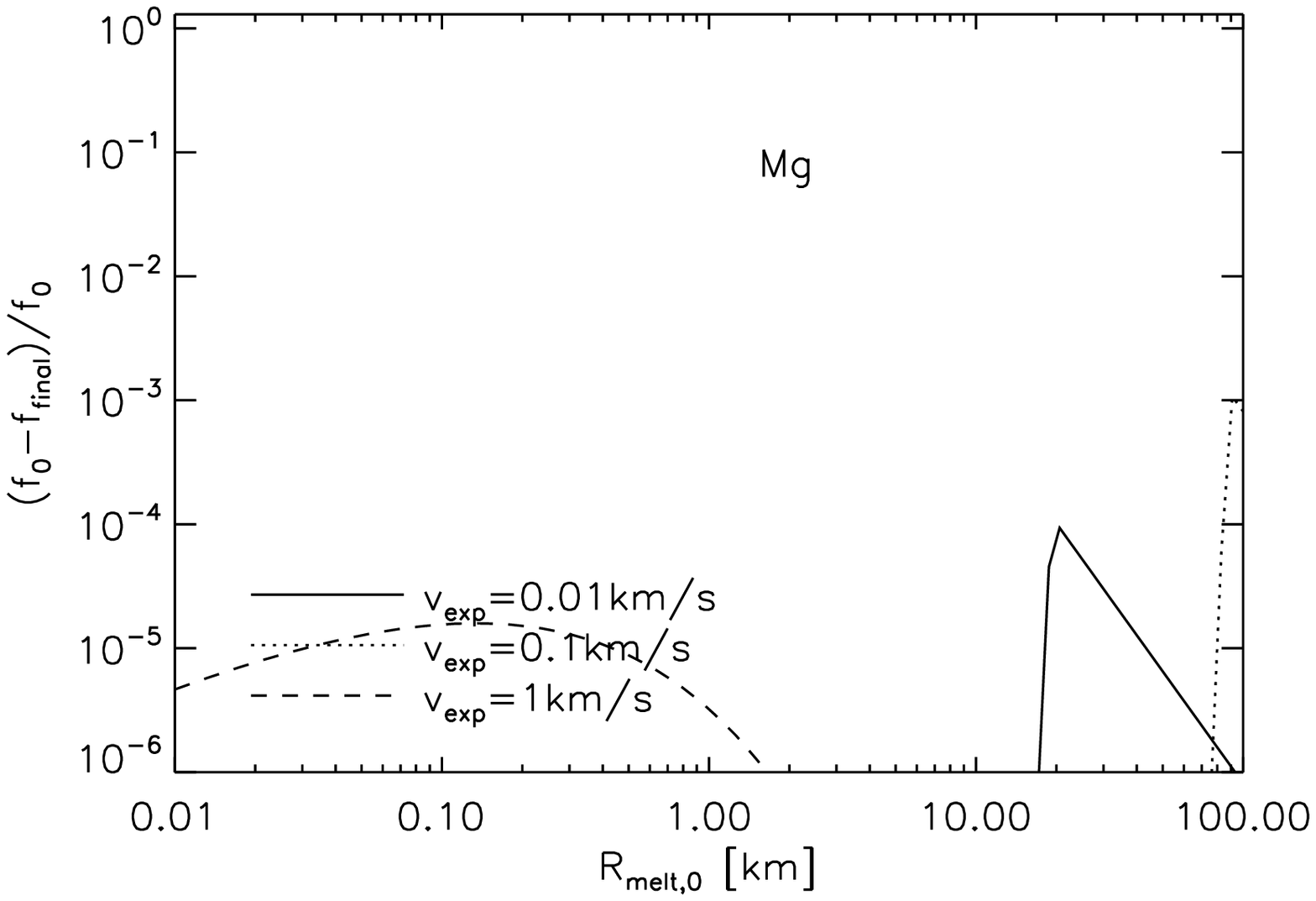}
\includegraphics[width=0.48\textwidth]{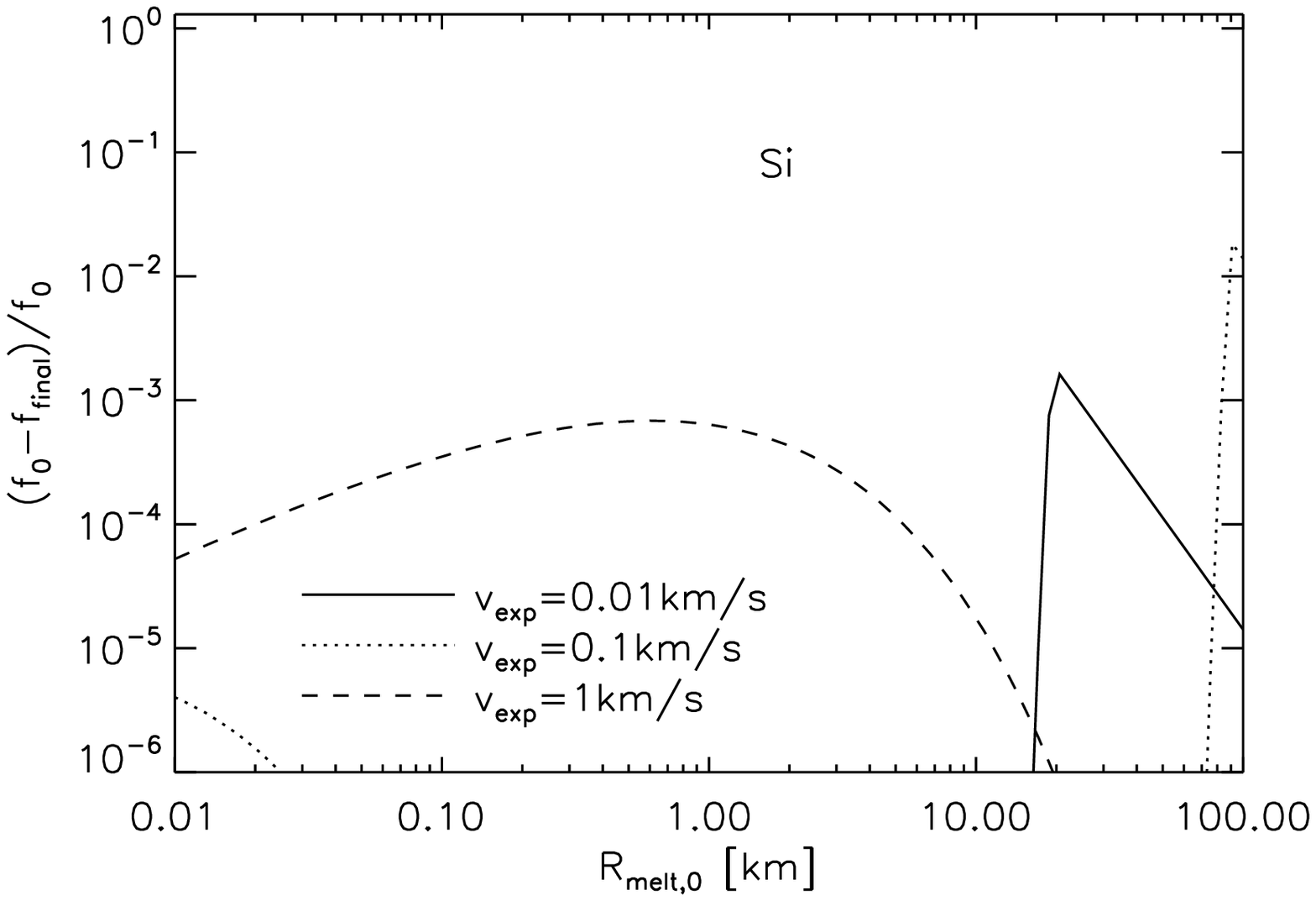}
\includegraphics[width=0.48\textwidth]{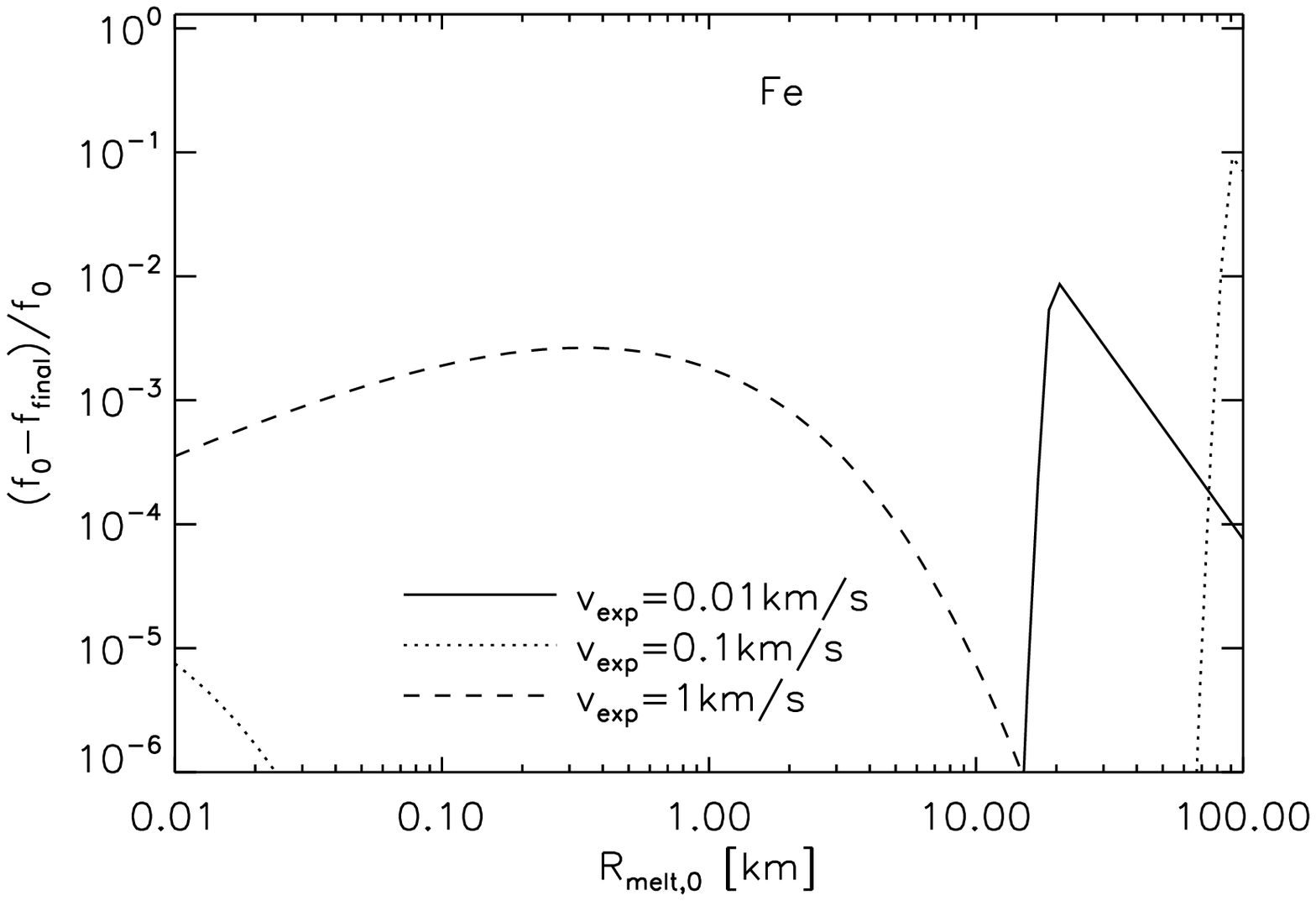}
\includegraphics[width=0.48\textwidth]{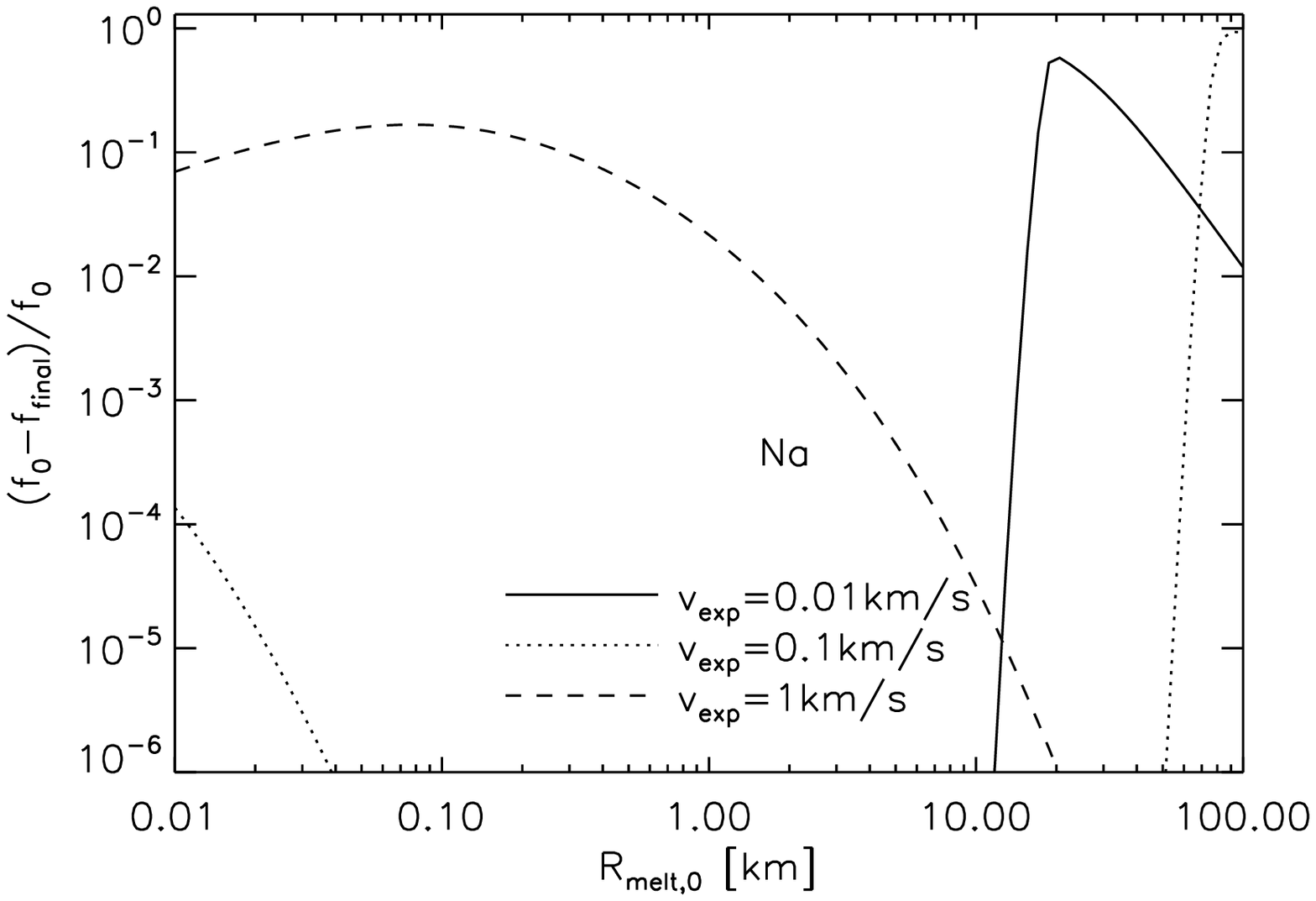}
\includegraphics[width=0.48\textwidth]{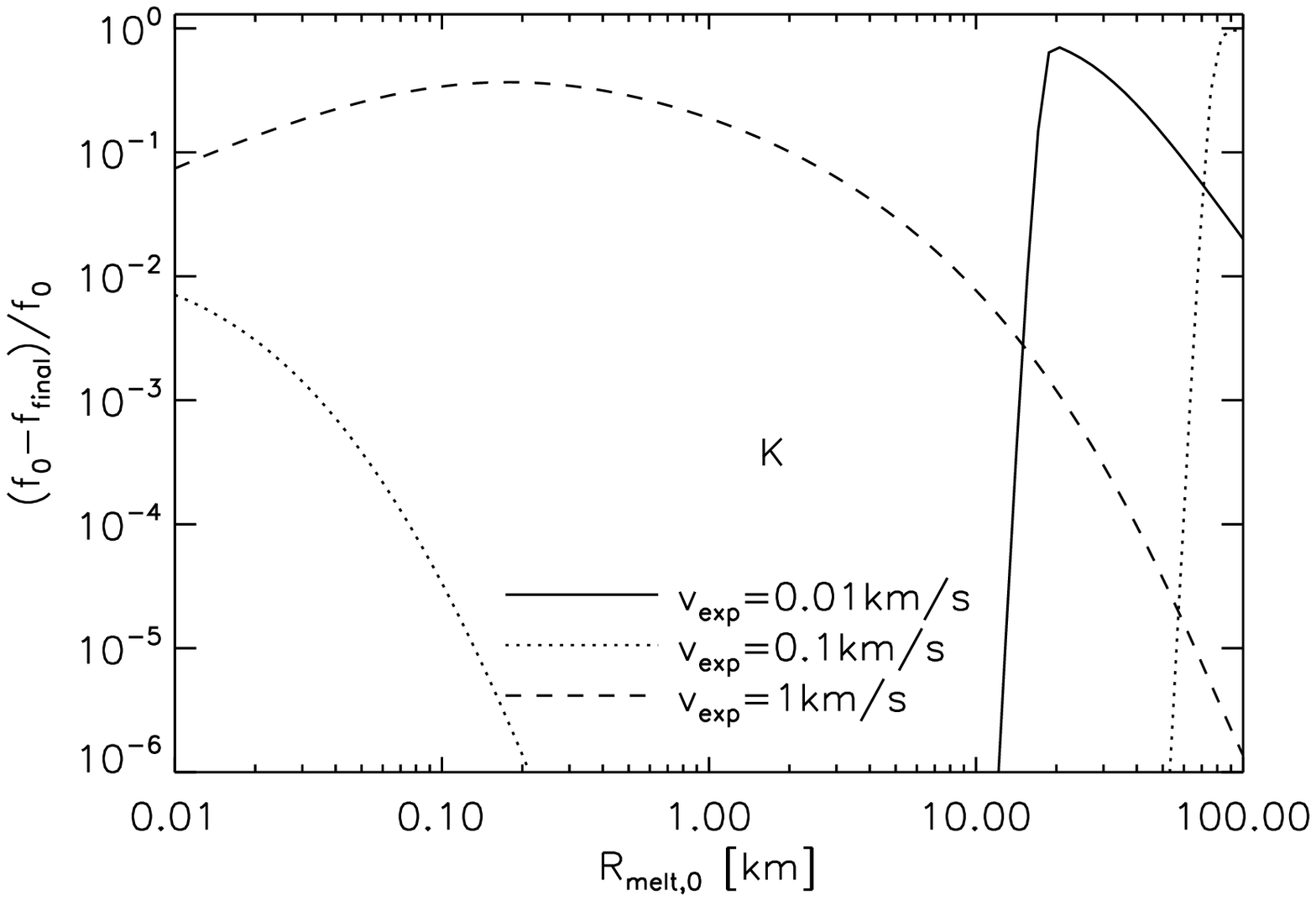}
\end{center}
\caption{\label{fig-volatile-loss-parscan}The fraction of volatiles that are
  lost from the droplets at the end of the expansion of the cloud. Plotted
  is $(f_0-f(t\rightarrow\infty))/f_0$ \modif{(where $f_0$ is the initial fraction
  of volatiles in the melt)} as a function of the initial mass of the cloud
  expressed in terms of the equivalent radius $R_{\mathrm{melt,0}}$
  (cf.~Eq.~\ref{eq-def-rmelt0}) for three values of the expansion velocity
  $v_{\mathrm{exp}}$. The analytic temperature model was used. }
\end{figure*}

The results of fiducial models F1 and F2 \modif{(see table \ref{tab-model-param})} are shown in
Figs.~\ref{fig-abun-feq-fvac}, \ref{fig-evapcond-timescales}. In Fig.\
\ref{fig-abun-an-num} the time-dependent abundance of all volatiles are
shown for models F1, F2, F3 and F4. The results show that at very early
times ($t\ll t_{\mathrm{cool}}$, \modif{well before the cloud temperature starts
to drop}) the evaporation of all volatiles nearly
instantly reaches saturation at the abundance $f_i=f_{\mathrm{eq},i}$ given by
Eq.~(\ref{eq-fequil}). This nearly instant saturation can be seen in
Fig.~\ref{fig-evapcond-timescales} which compares the various time scales
with each other for sodium. For early times the equilibration time scale
$t_{\mathrm{eq}}$ is clearly much smaller than $t$, so that the equilibrium
vapor pressure (and thus the equilibrium abundance) is easily reached within
that time. At these early times \modif{the corresponding abundance of the volatile
inside the lava droplet is nearly identical to the initial abundance} 
($f_{\mathrm{eq},i}\simeq f_{0,i}$) because the
volume per chondrule $V$ is still so small that there is no
space to deposit much vapor. In other words: at these early stages hardly
any volatile is in vapor form.  This can also be seen by looking at the
$t_{1\%}$ curve in Fig.~\ref{fig-evapcond-timescales}: only at a relatively
late time (10 minutes for model F1 and 0.1 minutes for model F2) will
$t_{1\%}$ become smaller than $t$, or in other words: before $t_{1\%}$ less
than 1\% of the volatile is in vapor form.

As time progresses and the volume of space around each chondrule increases,
the volatile abundance $f(t)$ inside the chondrules \modif{decreases} due to
evaporation. This is best studied by first looking at the equilibrium
abundance $f_{\mathrm{eq},i}(t)$ as a function of time, as shown in
Fig.~\ref{fig-abun-feq-fvac} for sodium. It shows that $f_{\mathrm{eq},i}$
drops more and more \modif{during the initial constant-temperature phase
  which lasts} until $t=t_{\mathrm{cool}}$, but \modif{once the cloud starts
  to cool down (for $t>t_{\mathrm{cool}}$)} it rises again. This is because
at low temperatures the vapor wants to recondense onto the chondrules, in
spite of the increase of volume per chondrule $V$. The question is: will the
actual abundance $f(t)$ follow the equilibrium abundance
$f_{\mathrm{eq},i}(t)$?  \modif{In other words: can the
  evaporation/condensation process in this model be approximated by a
  time-sequence of equilibrium states at ever increasing volume $V$ and
  decreasing temperature $T$?} This can be answered by looking again at
Fig.~\ref{fig-evapcond-timescales}. In particular in model F1 one sees that
the equilibration time scale $t_{\mathrm{eq}}$ for sodium is very much
smaller than $t$ until well after $t_{\mathrm{cool}}$. The abundance $f(t)$
thus has no problem following $f_{\mathrm{eq}}(t)$. This explains why in
Fig.~\ref{fig-abun-feq-fvac} for model F1 one cannot even distinguish
between the curves for $f(t)$ and $f_{\mathrm{eq}}(t)$.  This means that
while some of the sodium is turned into vapor around $t\simeq
t_{\mathrm{cool}}$, the rapid drop in temperature for $t> t_{\mathrm{cool}}$
will cause nearly all of the vapor to recondense back into the droplets. For
model F2 (high expansion velocity), however, the equilibration time scale
for sodium exceeds $t$ already soon after $t_{\mathrm{cool}}$, and it can be
seen in Fig.~\ref{fig-abun-feq-fvac} that $f_i(t)$ and
$f_{\mathrm{eq},i}(t)$ start to deviate from each other. The recondensation
does not manage to finish before the volume $V$ becomes so big that the
recondensation stagnates. This means that some sodium vapor remains in the
vapor phase and does not recondense, and thus some sodium is lost
permanently from the chondrules for model F2. \modif{This behavior can not
  be modeled by a sequence of equilibrium models, and is a clear case of
  time-dependent non-equilibrium evaporation/condensation. One is now
  compelled to ask: what will happen to this non-condensed vapor?  This is
  not trivial to answer and will depend on when the simple ballistically
  expanding cloud model will break down. A reasonable scenario is that once
  the chondrules and the remaining vapor dissipate into the solar nebula,
  the vapor will condense out onto the fine-grained dust in the nebula. In
  other words: the chondrules would then be somewhat depleted in these
  volatiles, while the fine-grained nebular dust (which is perhaps the
  future matrix?) will be enhanced in these volatiles.}

Fig.~\ref{fig-abun-an-num} shows the abundances of all five volatiles as a
function of time for all models. Both the full radiative transfer
temperature model as well as the analytic temperature model of
Eqs.~(\ref{eq-t-analytic},\ref{eq-t-analytic-zero}) are shown, and they
agree very well. In model F1, the model with an intermediate expansion
velocity, all volatiles recondense back onto the chondrules. We thus expect
that for these parameters no volatile losses should be found in chondrules
(modulo possible hydrodynamic escape effects). For model F2, the model with
a high expansion velocity (1 km/s), the volatile losses are considerable, at
least for Na and K. The elements Mg and Si seem to be entirely immune to
evaporation under these conditions, while Fe does appear to be slightly
affected for model F2.

The models show that if it were not for the high density of the cloud, the
sodium and potassium would have been almost completely lost from the
chondrules. That vapor would then have dispersed into the nebula and much
later would recondensed onto the fine-grained nebular dust. The ``savior''
of the volatiles is the fact that the expanding cloud of lava droplets
reaches its cooling time $t=t_{\mathrm{cool}}$ typically before the moment
when equilibration time $t_{\mathrm{eq},i}$ drops below $t$. That means that
if some volatiles may have vaporized during the expansion, the temperature
will drop soon enough for the vapor to recondense onto the chondrules before
the vapor would be lost.  Moreover, for Fe, SiO and Mg most of the volatile
elements never even reach the gas phase because before $t=t_{\mathrm{cool}}$
the available volume per chondrule is still too small to be able to store
much of the volatile element in vapor form. Only for Na and K there is a
moment (around $t=t_{\mathrm{cool}}$) where most of the volatile is in vapor
form. However, soon after $t_{\mathrm{cool}}$, as the temperature drops,
some (model F2) or all (model F1) of the vapor recondenses again. 

Model F3 is a model of a very massive and slowly expanding cloud.
Interestingly for this model the volatile abundances reach the lowest
intermediate values of all models before returning completely back to their
original values. This may appear counter-intuitive, as we expect the slow
expansion to keep the volume around each chondrule small for a long
time. The essential time scale, however, is the time at which the cooling
starts. If the volume around each chondrule reaches large enough values for
most of the volatile to escape the lava droplets before the cooling sets in,
then the depletion at $t=t_{\mathrm{cool}}$ will be the largest.

Model F4 is the most extreme one, where the cloud expands so fast that any
volatiles lost until $t=t_{\mathrm{cool}}$ will not have the time to
recondense before the cloud has essentially dispersed. The lost vapor will
then presumably recondense out on nebular dust grains.

The results of all models do depend somewhat on position within the
cloud. The results shown so far are for the center of the cloud. Toward the
edge of the cloud the cooling sets in earlier, and hence also the
recondensation.  Fig.~\ref{fig-abun-na-afo-r} shows, as an example, the
radial dependence within the cloud of the Na-abundance in model F1 for
several time snapshots.  As radial coordinate the dimensionless
$\eta=r/R_{\mathrm{cloud}}$ is used, so that the edge of the cloud is, at
each time snapshot, located at the same position $\eta=1$. In other words:
that the results are all shown at the same relative spatial scales. One sees
that the strongest depletion occurs at intermediate times in the center of
the cloud, whereas toward the edge less depletion occurs. This is because
the outer regions of the cloud are cooler, and therefore the equilibrium
vapor pressure is lower.

It is interesting to investigate how much of the vapor still recondenses
after the solidus temperature is reached. This vapor would then produce a
layer of high concentration of the element around the chondrule. Any vapor
that recondenses shortly before the solidus is reached may still partly
diffuse into the chondrule but perhaps not perfectly, predicting that that
volatile element would have a higher concentration near the surface of the
chondrule than near the center. The time of reaching the solidus temperature
is marked in Fig.~\ref{fig-abun-an-num}.

Fig.~\ref{fig-volatile-loss-parscan} shows the final results of the models:
the fraction of volatiles permanently lost from the chondrules at the end of
the models for various cloud masses and expansion velocities. This shows
that strong losses are only expected for extremely high expansion velocities
($\gtrsim $1 km/s), or for very massive ($R_{\mathrm{melt,0}}\gtrsim 10$ km)
but slowly expanding clouds. This opens the possibility that measurements of
the degree of volatile loss can distinguish between low and high velocity
impacts and/or large or small vapor cloud masses. All these results are,
however still without taking into account the hydrodynamic loss of the
volatile.

\subsection{Justification for the simplified evaporation/condensation model}
\label{sec-vol-checksimplemodel}
As we described in Section \ref{sec-simplified-model-evapcond}, we
introduced several approximations in the evaporation/condensation model, in
order to keep the model simple and easily reproducable. However, it is fair
to ask to which extent these simplifications may affect the overall results.
The two most serious approximations we made were (1) that
$p^{\mathrm{eq}}_i\propto f_i$ and (2) that $p^{\mathrm{eq}}_i$ does not
depend on $f_{k\neq i}$. The evaporation/condensation theory reviewed in the
appendices shows that neither are really correct. For Na and K we expect,
from theory, more something like $p^{\mathrm{eq}}_i\propto \sqrt{f_i}$, and
the vapor pressures do influence each other. 

\begin{figure*}
\begin{center}
\includegraphics[width=0.48\textwidth]{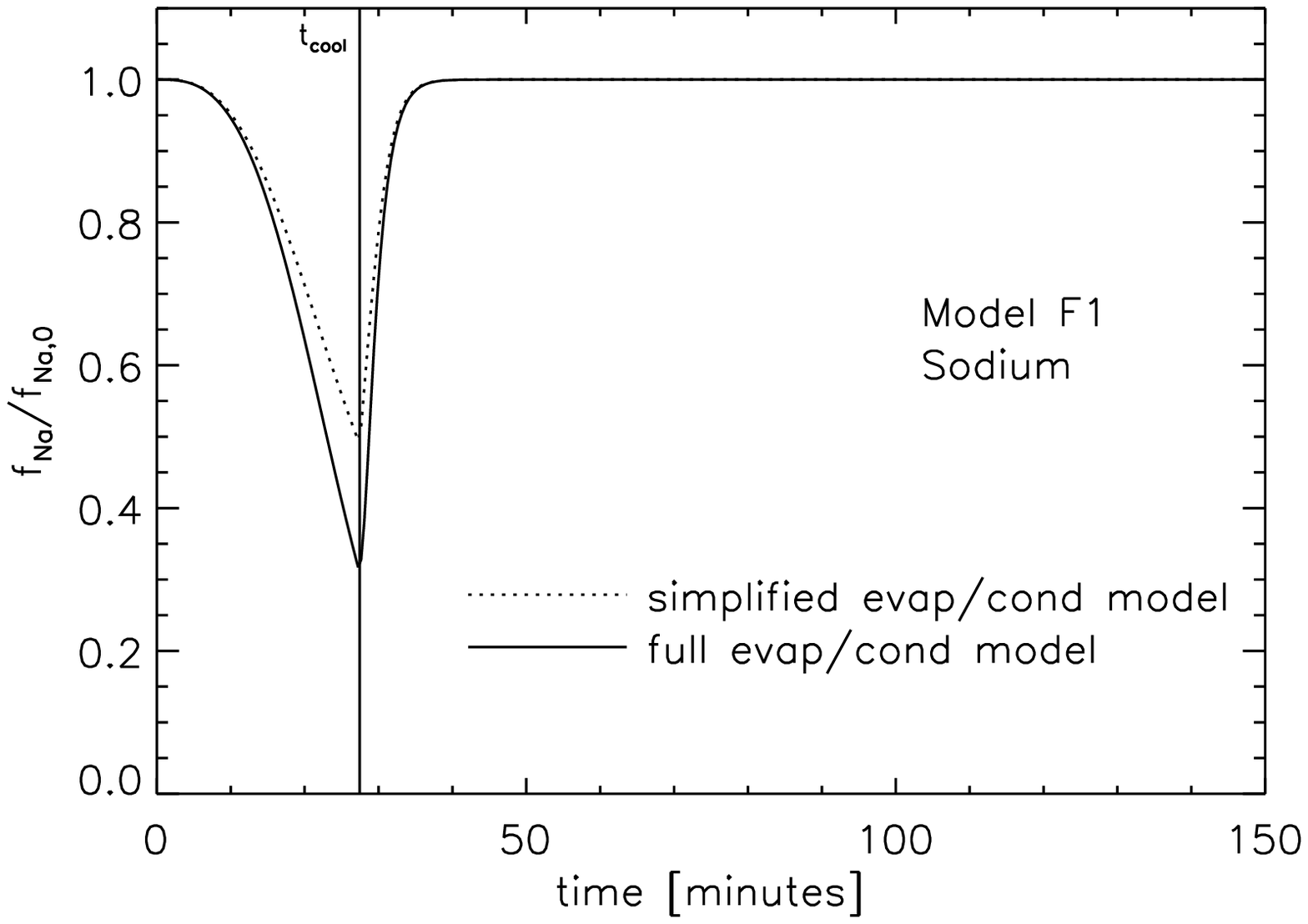}
\includegraphics[width=0.48\textwidth]{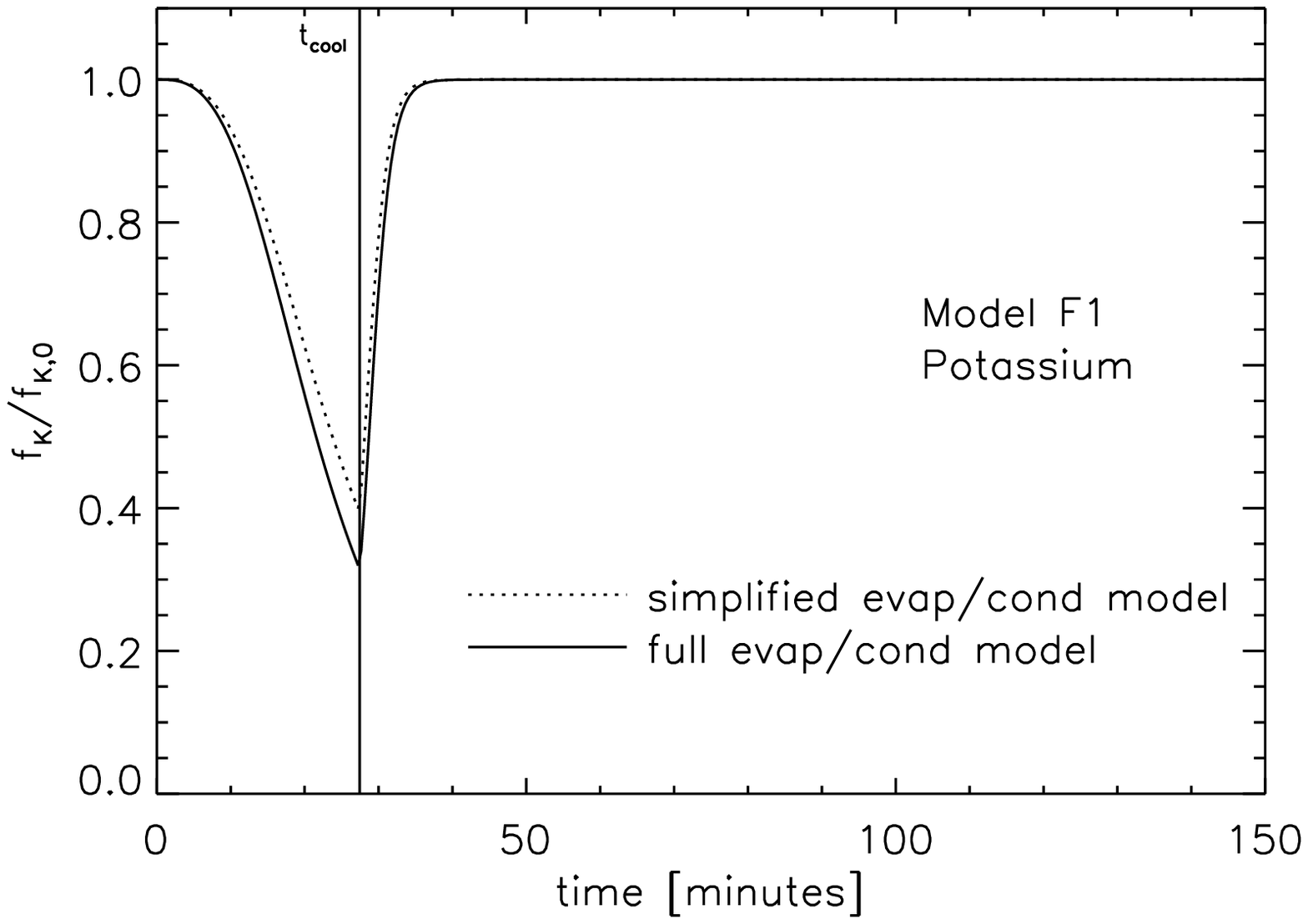}
\end{center}
\begin{center}
\includegraphics[width=0.48\textwidth]{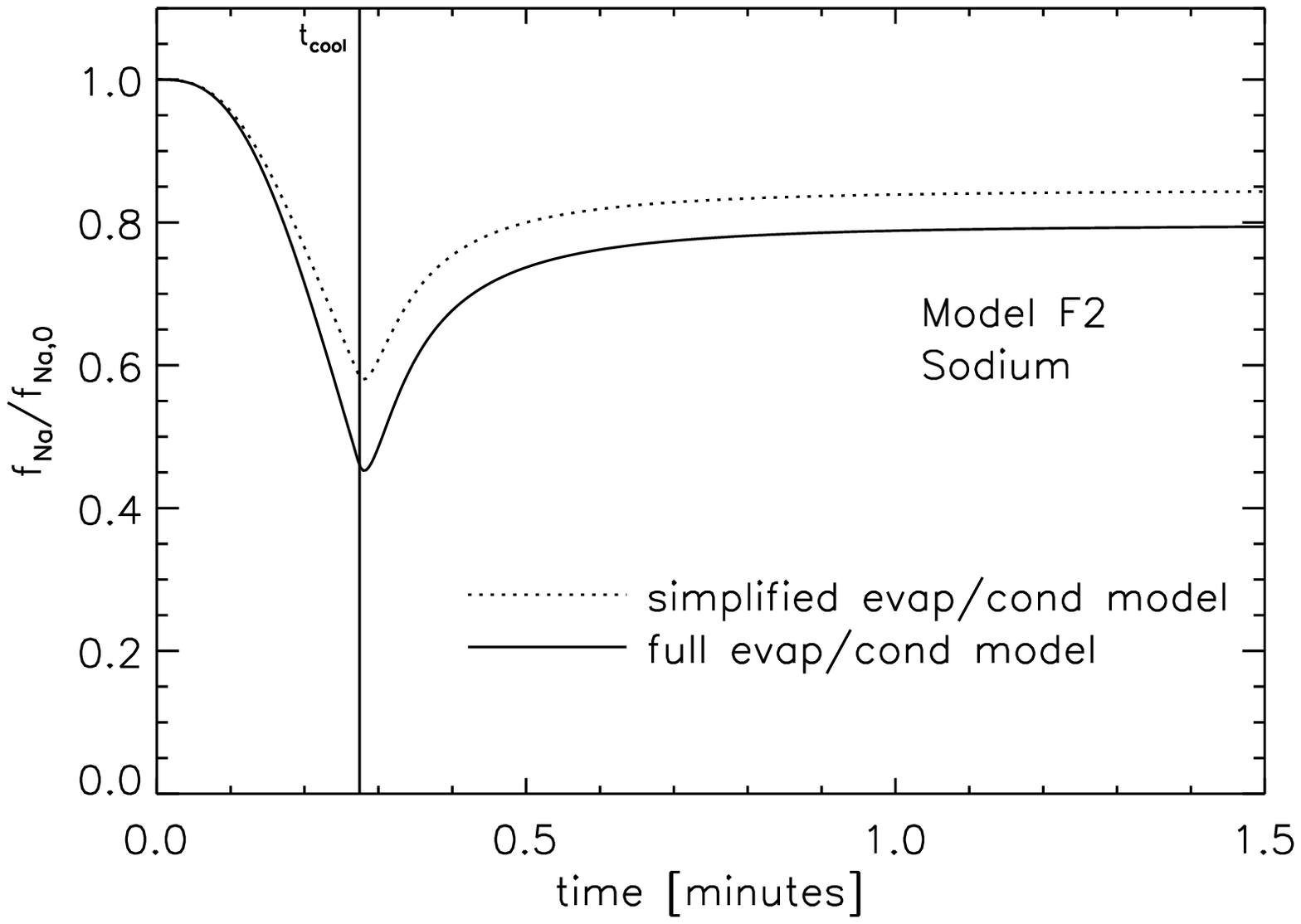}
\includegraphics[width=0.48\textwidth]{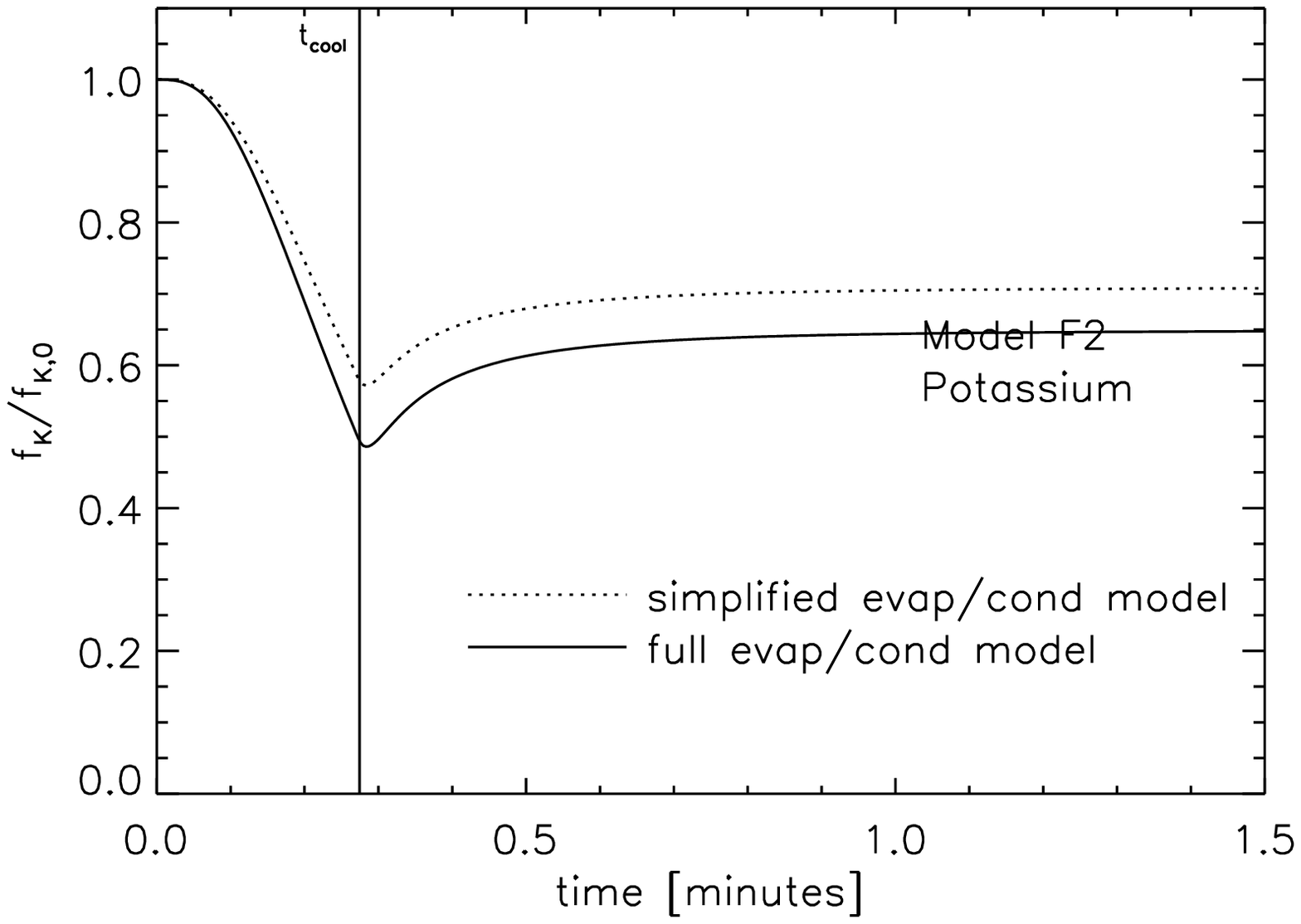}
\end{center}
\caption{\label{fig-justify-simple}Comparison of the standard model with our
  simplified description of evaporation and condensation (dotted line) to a
  version of the model in which the equilibrium pressures are recalculated
  at every time step using the full machinery of evaporation/condensation
  described in the appendices (solid line). Top panels: Model F1 ($R_{\mathrm{melt,0}}=1\,\mathrm{km}$, 
  $v_{\mathrm{exp}}=0.1\mathrm{km/s}$), bottom
  panels: Model F2 ($R_{\mathrm{melt,0}}=0.1\,\mathrm{km}$, 
  $v_{\mathrm{exp}}=1\mathrm{km/s}$). Left panels: sodium, right panels: potassium.}
\end{figure*}

In order to test whether these simplifications strongly affect the result
(and thus make the result unreliable) we created also a model computer code
where the equilibrium vapor pressures $p^{\mathrm{eq}}_i$ are recomputed at
every time step using the full machinery described in the appendices. The
only things we still keep constant are the activity coefficients, because
they follow from the MELTS code, for which we do not have a version
available at present that can be directly linked into our program. We
compare the results for Na and K, for models F1 and F2. The results are
shown in Fig.~\ref{fig-justify-simple}.

The results show that the simplified evaporation / condensation model does a
very good job for Na and K compared to the more sophisticated model. The
differences are presumably substantially smaller than many of the other
model uncertainties, such as initial condition parameters. The simplified
model is therefore, for these purposes, good enough.

\subsection{High-temperature models}
\label{sec-high-temp-models}
\begin{figure*}
\begin{center}
\includegraphics[width=0.48\textwidth]{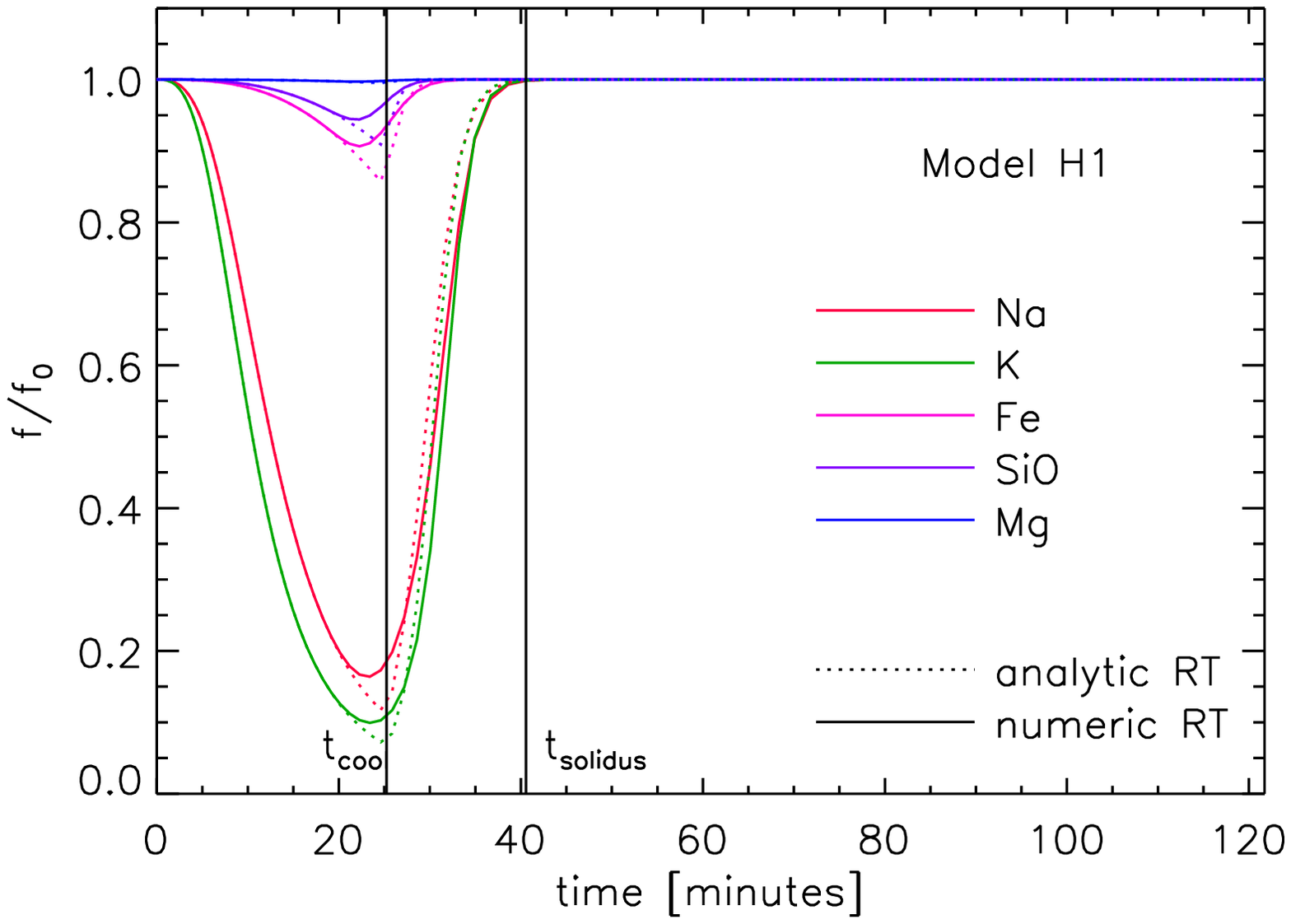}
\includegraphics[width=0.48\textwidth]{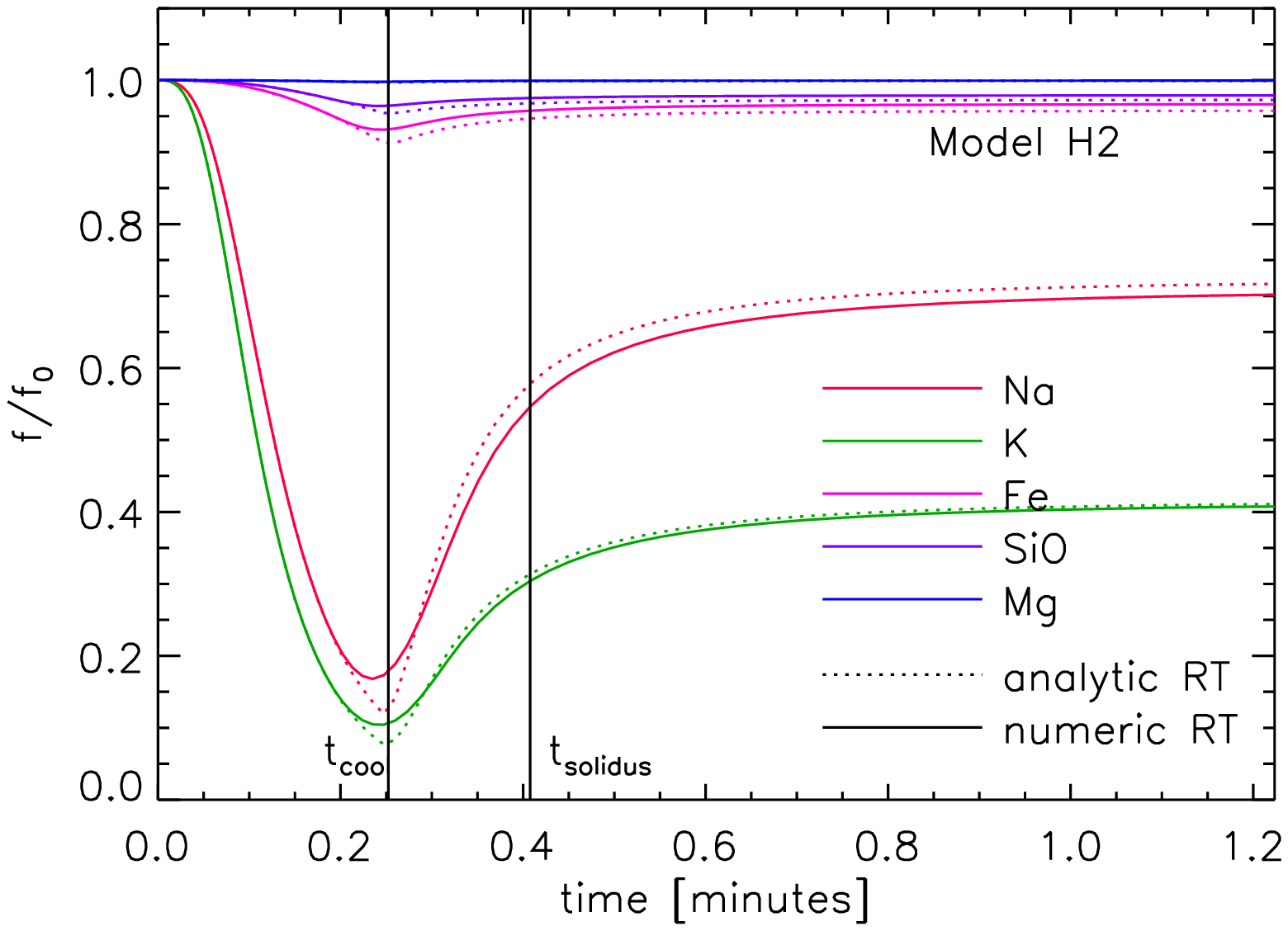}
\end{center}
\begin{center}
\includegraphics[width=0.48\textwidth]{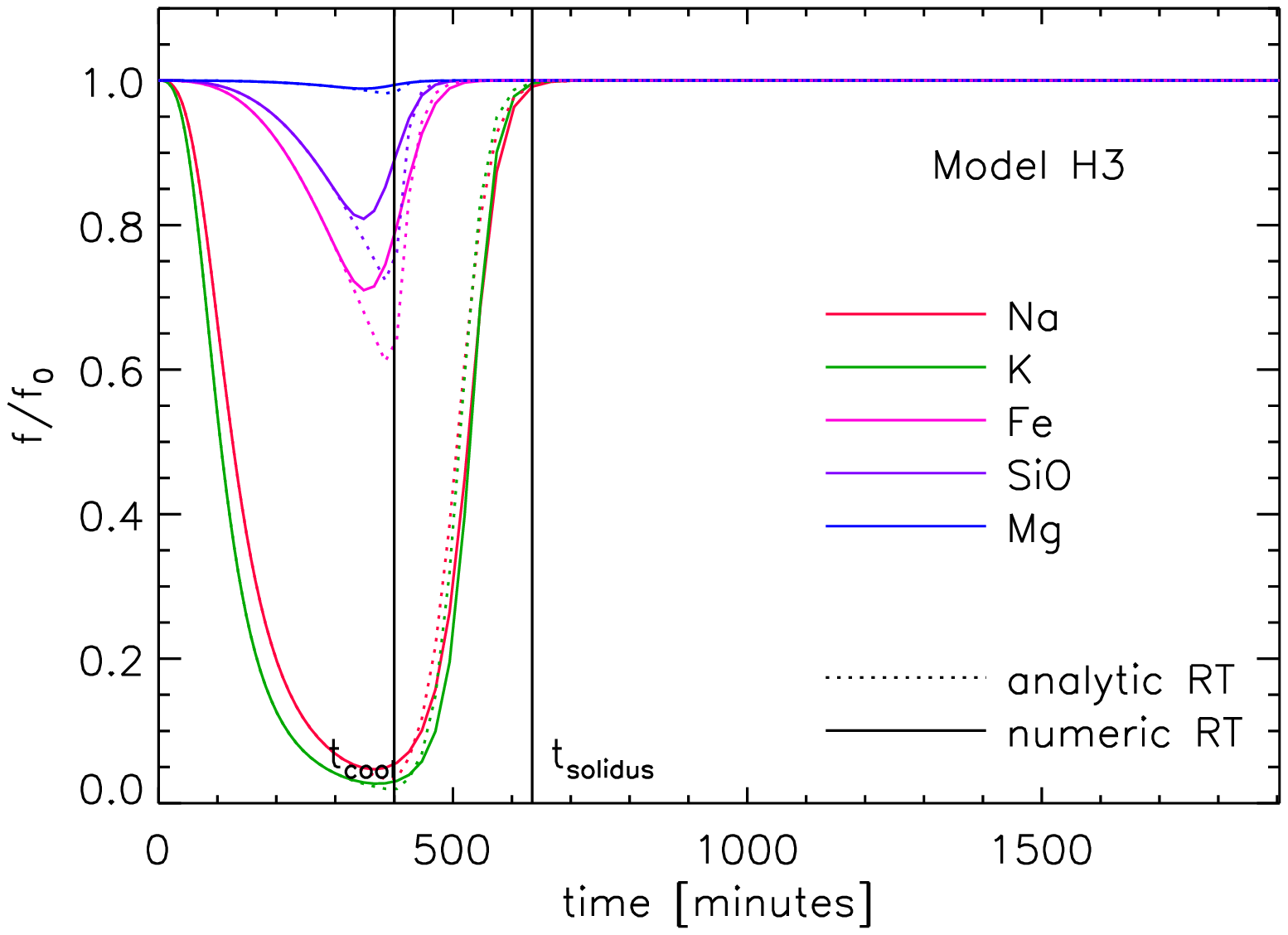}
\includegraphics[width=0.48\textwidth]{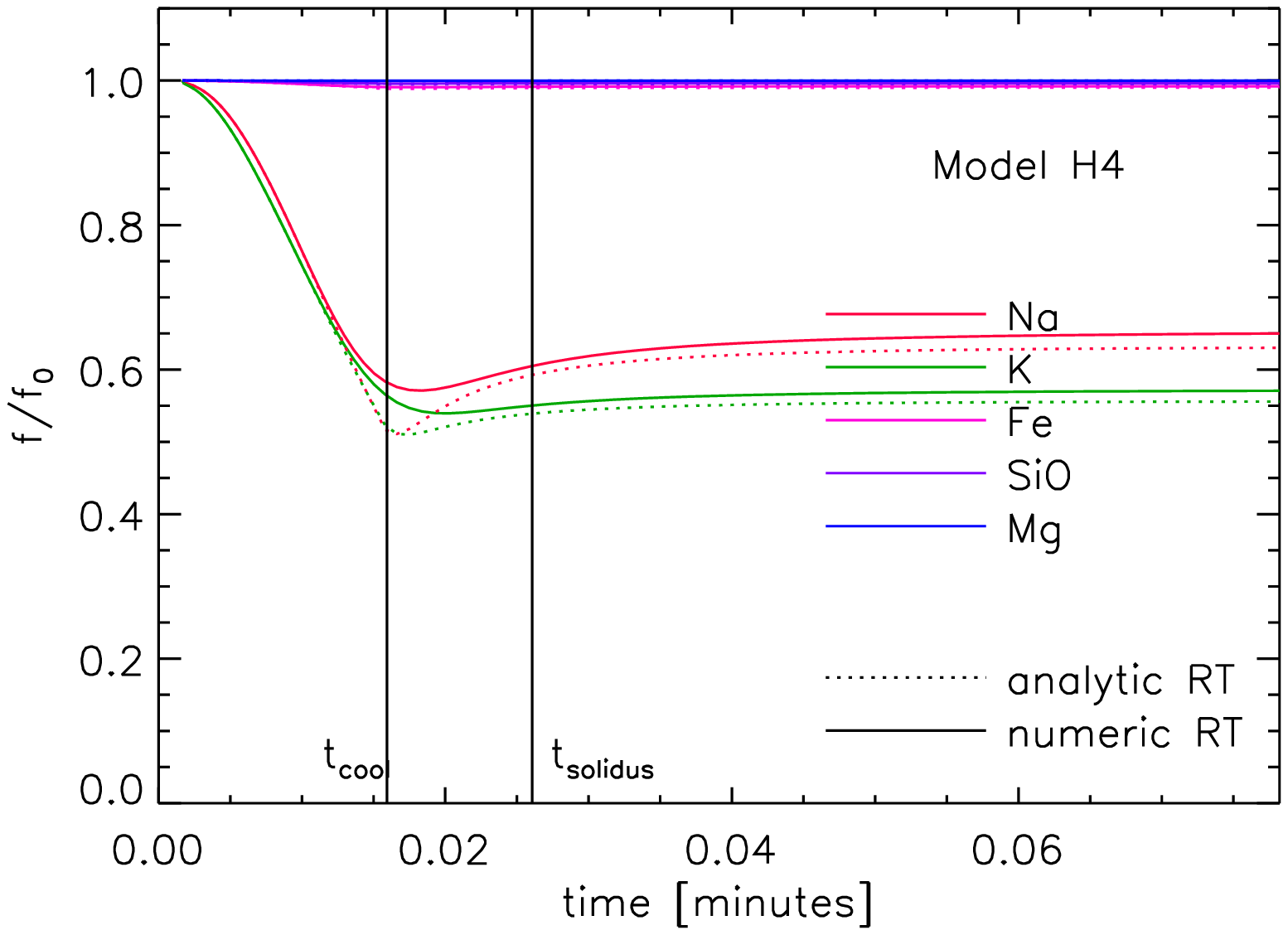}
\end{center}
\caption{\label{fig-abun-hightemp}\modif{As Fig.~\ref{fig-abun-an-num}, but now for
  a higher initial temperature: $T=2300$ K. These models are marked as H1,
  H2, H3 and H4. For the rest the models are the same as the models F1, F2,
  F3 and F4 (Fig.~\ref{fig-abun-an-num}).}}
\end{figure*}

\modif{The fiducial models F1, $\cdots$, F4 all start with a temperature of
$T=2000$ K. But what happens if the initial temperature is considerably
higher? We reran all four models with an initial temperature of $T=2300$
K. We name these models H1, $\cdots$, H4. The results are shown in
Fig.~\ref{fig-abun-hightemp}. It is seen that the main consequence is that
the Na and K reach lower minima, and at the end more Na and K remains lost
(in particular in models H2 and H4).  The other elements do not experience
much depletion at the end, but in models H1 and H3 the Fe and SiO are
temporarily reduced after which they recondense again.}

\modif{In type IAB chondrules one often finds low-Ca pyroxene rims (e.g.~Hewins \&
Zanda 2012) which are interpreted as evidence for evaporation and
recondensation of SiO. In the fiducial models (Fig.~\ref{fig-abun-an-num})
the temperature is too low for appreciable SiO evaporation. But in the high
temperature models (except for H4) a substantial evaporation and subsequent
recondensation of SiO is observed, which might be consistent with such
low-CA pyroxene rims, since the recondensing SiO may not have time to mix
well with the cooling droplet. Fe evaporation usually accompanies SiO
evaporation, but in our model we include only a single oxidation state of
iron, so it would be premature to draw too strong conclusions from this.}

\modif{It should be kept in mind that in these high temperature models (in
particular in model H3) the original assumption that the radius of the
chondrules does not change is no longer correct. With half the loss of mass,
the radius is 20\% smaller. Therefore these model results merely give a
qualitative picture of what happens.}

\section{Discussion and conclusion}
\label{sec-discussion}
In this paper we computed the degree of loss of volatile elements from
chondrules formed in an impact-splash scenario. We use the simple
spherically symmetric expanding chondrule cloud model of Paper I which gives
us the chondrule density and temperature of the cloud as a function of
radial coordinate inside the cloud and as a function of time. That model
showed that initially the cloud stays at a constant temperature, but after
a well-defined time starts to cool down due to radiative cooling. On the
basis of this model we calculate how the volatile compounds Na$_2$O, K$_2$O,
FeO, SiO$_2$ and MgO evaporate out of the molten chondrules but quickly
reach their vapor equilibrium pressure, effectively halting the further
evaporation. We find that shortly before the cloud starts to cool some
elements (mainly Na and K) start to experience non-negligible losses from
the chondrules (tens of percents). However, as the cooling sets in these
elements usually rapidly recondense onto the chondrules.

The models of this paper show that volatile retention appears to be a
natural consequence of the impact-splash origin scenario for
chondrules. Since volatile retention is one of the critical observables in
chondrules, our model calculations provide some further support for the
impact splash scenario. 

Depending on the parameters of the model (the cloud mass and expansion
velocity) the recondensation can \modif{also, however, be incomplete (see
  e.g.~model F2)}, leaving some signature of volatile loss in the
chondrules. The fate of the remaining vapor is likely to recondense at a
later stage onto the dust grains of the solar nebula. \modif{However, this
  typically happens only in the models with very rapid cooling (for model F2
  this is 15 seconds), which appears to be inconsistent with typical
  chondrule textures. The models with many hours of cooling time appear to
  have complete recondensation. Since there exist in Nature chondrules with
  substantial Na and K depletion, as well as some chondrules with enhanced
  abundances, we speculate that more detailed physics of the expanding cloud
  has to be included, such as hydrodynamic flow of Na and K vapor through
  the cloud. Indeed,} the present model assumes that the vapor stays inside
the cloud, i.e.\ that there is no hydrodynamic escape. Also it is assumed
that no nebular gas permeates into the cloud, at least not before the
solidus temperature is reached. A future study will have to clarify if and
how the results are affected by this aspect. One possible outcome could be
that in addition to depleted abundances, the outer cloud regions may
experience a surplus of volatiles because when volatiles from the cloud
interior flow into the cooler cloud edge this could lead to accumulation of
these elements. 

Another issue that requires further study is whether all the recondensed
vapor can quickly mix throughout the chondrule, to yield a homogeneous
abundance, or whether the surface of the condrules are expected to be
enriched in volatile elements that recondensed in the late phases of the
present model.

\modif{In spite of its simplicity, the predictions of our model appear to be
  consistent with the conclusions of Hewins, Zanda \& Bendersky (2012) who
  find that there is evidence from Semarkona type II chondrules of
  evaporation and later recondensation of Na into these chondrules. This is
  indeed what our models show (see e.g.~Fig.\ref{fig-abun-an-num}). These
  authors also write that ``Type IIA chondrules lost more than half their Na
  and recondensation was incomplete'' (consistent with our models F3 and
  F4), while ``Type IIAB chondrules recovered most of theirs in their
  mesostases'' (consistent with our models F1 and F2). In their section 4.3
  they write ``Even though the high concentrations of moderately volatile
  elements indicate that Type II chondrules may have formed in a high
  density region, there is still evidence of at least partially open system
  behavior.'', which is indeed what our model finds if the expansion and
  cooling is too quick for all the Na to recondense (models F3 and F4),
  thought that may require too fast cooling to be consistent with 
  textures.}

\modif{Type I chondrules have low Fe and low Na content. Our model shows
  that Fe does not appreciably evaporate during the splash process, while Na
  does. It therefore seems that the low content of both elements in type I
  chondrules should be be related to another process.}

\modif{Type IAB chondrules show evidence of recondensation of SiO, which
in our model would be consistent with higher initial temperatures.}

\modif{Finally, it has long been thought that the chondrules from type CB
  (Bencubbin-like) chondrites originate from a hypervelocity impact
  (e.g.~Krot et al.~2005). In addition to chondrules, these chondrites
  contain large amounts of metal spherules thought to have condensed out of
  the vapor phase. In fact, the Fe-Ni metal dominates by $\sim$ 60 \% the
  volume of these chondrites, and it is thought that the chondrules and the
  metal spherules formed during the same impact. Our model as it stands is
  too simple to simulate the conditions of this highly energetic event with
  high metal content. But the fact that they are highly depleted in Na and K
  and that they have only non-porphyritic textures without relic grains or
  rims (see Krot et al.~2005), suggests, in the context of our model and
  consistent with conclusions by other authors, that these formed in a very
  high-speed impact. Comparing to our model this scenario suggests high
  initial temperature and a fairly small cloud (i.e.~small
  $M_{\mathrm{cloud}}$) with high plume expansion velocity
  $v_{\mathrm{exp}}$. The high temperatures are probably necessary to create
  the Fe vapor, and the small, rapidly expanding cloud parameters ensure
  that the cooling is rapid and the Na and K vapor do not recondense. Our
  models H2 and H4 might go somewhat in that direction, but are probably not
  extreme enough. If we would stretch our model parameters too far, however,
  the underlying assumptions are broken and the model would be no longer
  valid. Qualitatively, however, it seems that the origin of CB chondrites
  as a result of an impact, as suggested by many authors in the past, fits
  well into our scenario.}

In future studies it is important to study more realistic geometries. The
nice thing of our simple spherically symmetric model is that many things can
be quantitatively calculated quite easily. But it is clear that in reality
the splashes from impacts of pre-molten bodies are not spherically expanding
clouds. Smooth particle hydrodynamics (SPH) calculations may be required.

\modif{The evidence of chondrules being formed in high-pressure enviroments
  due to the retention of Na and K remains, however, subject to debate. Na
  and K may also re-enter at a much later stage, for instance due to contact
  with liquid water on the parent body, since Na and K are water-soluble.
  So it is important to study more detailed consequences of models of the
  kind presented in this paper. But the models so far seem to produce
  relatively robust results, i.e.~not too strongly dependent on
  ill-constrained parameters. The partial evaporation and evaporation with a
  retention of all or most of the volatile elements Na and K is a very
  natural outcome of the model and requires no fine-tuning. That gives some
  confidence that impact splashes (be they due to hypervelocity impacts with
  jetting or due to lower speed impacts of pre-heated bodies) may be a
  natural explanation of the formation of chondrules.}

{\bf Acknowledgements:} We would like to thank Andreas Pack, Eric Gaidos,
Knut Metzler, Alessandro Morbidelli, David Lundberg and Stephan Henke for
useful discussions and feedback. We also thank the anonymous second 
referee for very useful input and suggestions.

This work has been supported by the Deutsche Forschungsgemeinschaft
Schwerpunktprogramm (DFG SPP 1385) ‘‘The first 10 Million Years of the Solar
System -- a Planetary Materials Approach’’ (grants Du 414/12 and Du 414/14).

A.J. is grateful for the financial support from the European Research Council
(ERC Starting Grant 278675-PEBBLE2PLANET), the Knut and Alice Wallenberg
Foundation, and the Swedish Research Council (grant 2014-5775).

%

\appendix

\section{Equilibrium vapor pressures for metal oxide melts}
\label{sec-equil-pvap-metal-oxides}
For the calculation of time-dependent evaporation and condensation using the
Hertz-Knudsen equation we need to know the equilibrium vapor pressures of
the vapor species for a given temperature. These can be calculated from
thermodynamic properties measured in the laboratory. In our case we are
interested in evaporating molten rock. We will not consider the presence of
$H_2$ gas nor an externally fixed oxygen pressure, but instead assume that
the only gas that is present is the evaporated vapor itself. For our model
of an expanding cloud of lava droplets resulting from an impact this
assumption this is presumably reasonably valid. The procedure described here
can, however, be easily generalized to the case of the presence of
pre-existing gas. The procedure we discuss here is a simplified version of
the procedure described in Fedkin et al.~(2006).

Since the lava droplet is a mixture of different metal oxides, let us focus
on one of them (our `metal oxide of interest') with a mass fraction denoted
by $f$. As an example we choose Na$_2$O, but the following discussion will
also apply to other species. If the liquid would be an ideal fluid, one
could use Raoult's law which states that the equilibrium vapor pressure of
the species interest equals $f$ times the equilibrium vapor pressure of the
pure species: $p^{\mathrm{eq}}=f\,p^{\mathrm{eq,pure}}$. One can regard this
as saying that a fraction $f$ of the surface of the melt is made out of our
metal oxide of interest. This would greatly simplify the calculation of
evaporation because we can then treat the evaporation of each species
independent from the others.

Unfortunately for metal oxides the situation is more complex. First of all,
molten metal oxides are usually far from ideal fluids. The linear relation
$p^{\mathrm{eq}}=f\,p^{\mathrm{eq,pure}}$ does not necessarily hold. Instead,
we must use the activity $a$ rather than the mass fraction $f$, and the
equilibrium vapor pressure will not necessarily be linear in $a$ (nor in
$f$). Secondly, the evaporating metal oxides tend to dissociate into its
most stable gaseous atomic or molecular constituents. For instance, for
Na$_2$O we have the following evaporation/condensation reaction:
\begin{equation}\label{eq-na2o-2nao2}
\mathrm{Na}_2\mathrm{O} (l) \leftrightarrow 2\mathrm{Na} (g) + \tfrac{1}{2}\mathrm{O}_2 (g)
\end{equation}
where $(l)$ means liquid and $(g)$ means gas.  Note that in the melt the
metal can also be locked up in more complex forms. For Na this can be
e.g.\ Na$_2$SiO$_3$(l). The evaporation/condensation reaction then
becomes
\begin{equation}\label{eq-na2sio3-2nasio2o2}
\mathrm{Na}_2\mathrm{Si}\mathrm{O}_3 (l) \leftrightarrow 2\mathrm{Na} (g) + \mathrm{Si}\mathrm{O}_2 (l) + \tfrac{1}{2}\mathrm{O}_2 (g)
\end{equation}
In either case there are more than one gas species coming out of the liquid,
typically one being the metal and the other being oxygen. This means that in
addition to the metal vapor pressure (in this case $p_{\mathrm{Na}}$) we
also have the oxygen vapor pressure $p_{\mathrm{O}_2}$ (which is equal to
the oxygen fugacity since the vapor can be treated as an ideal gas). The
metal equilibrium vapor pressure is then coupled to the oxygen vapor
pressure because for each metal vapor particle also an oxygen vapor particle
is ejected from the melt. Since all other evaporating metal species also
depend on oxygen, the oxygen vapor pressure couples all metal vapor
pressures to each other. 

The equilibrium vapor pressure for the metal of interest can be computed
from the equilibrium chemistry equation. Let us write the general form of
the evaporation reaction as
\begin{equation}\label{eq-metalox-evap-reaction-general}
\mathrm{M}_c\mathrm{X} (l) \leftrightarrow c\,\mathrm{M} (g) 
+ d\,\mathrm{Y} (l) + e\,\mathrm{O}_2 (g)
\end{equation}
where `M' is the metal, `X' is the combination of oxygen and possibly other
elements that together with `M' make up the compound, `Y' is what remains of
the liquid after the evaporation. If no remaining liquid substance `Y' is
involved we set $d=0$. For reaction (\ref{eq-na2o-2nao2}) we would have
M=`Na', X=`O', $c=2$, $d=0$ and $e=0.5$. For reaction
(\ref{eq-na2sio3-2nasio2o2}) we have M=`Na', X=`SiO$_3$', $c=2$,
$d=1$, Y=`SiO$_2$' and $e=0.5$.  The equilibrium partial pressure
$p_{\mathrm{M}}^{\mathrm{eq}}$ (in units of bar) of metal vapor species
$\mathrm{M} (g)$ can be calculated using the following equation:
\begin{equation}\label{eq-pvap-in-kequil}
K_{\mathrm{evap}} = \frac{{p_{\mathrm{M}}}^c{p_{\mathrm{O}_2}}^e{a_{\mathrm{Y}}}^d}{a_{\mathrm{M}_c\mathrm{X}}}
\end{equation}
where $K_{\mathrm{evap}}$ is the equilibrium constant for the evaporation
reaction, $p_{\mathrm{M}}\equiv p_{\mathrm{M}}^{\mathrm{eq}}$ is the metal
gas partial pressure (in units of bar), $p_{\mathrm{O}_2}$ is the oxygen gas
partial pressure (in units of bar), $a_{\mathrm{M}_c\mathrm{X}}$ the
activity of our metal oxide of interest in the fluid and (if present)
$a_{\mathrm{Y}}$ the activity of the remaining liquid substance Y. Note that
for convenience we will usually simply write $p_{\mathrm{M}}$ where actually
the equilibrium vapor pressure $p_{\mathrm{M}}^{\mathrm{eq}}$ is meant. The
difference becomes only relevant in non-equilibrium cases. Note also that
the reaction Eq.~(\ref{eq-metalox-evap-reaction-general}) is not the most
general one (for instance the evaporation of SiO$_2(l)$ into SiO$(g)$
+$\tfrac{1}{2}$O$_2(g)$ is not strictly included, but the principle stays
the same).

The equilibrium coefficient $K_{\mathrm{evap}}$ of the evaporation reaction can
be computed by
\begin{equation}\label{eq-equilconst-gibbs}
K_{\mathrm{evap}} = \exp\left(-\frac{\Delta_rG(T)}{RT}\right)
\end{equation}
where $R$ is the gas constant, $T$ the temperature, $\Delta_rG(T)$ is the
change of Gibbs free energy for the reaction and is given by
\begin{equation}\label{eq-deltarg}
\Delta_rG(T) = c\,\Delta_fG_{\mathrm{M}}(T) + e\,\Delta_fG_{\mathrm{O}_2}(T) 
+ d\,\Delta_fG_{\mathrm{Y}}(T) - \Delta_fG_{\mathrm{M}_c\mathrm{X}}(T)
\end{equation}
where for each species the $\Delta_fG(T)$ is the amount of Gibbs free energy
needed to create that species from its elements in their most stable phase
at that temperature (the reference phase). In appendix \ref{app-janafberman}
the values of $\Delta_fG(T)$ for all the relevant species of this paper
are given in the form of a polynomial fit to the laboratory data in the
temperature range between 1400 and 2200 K.

We assume in our model that the only gas that is present is the vapor
itself. If a single metal vapor species is dominating the oxygen production,
then there is a simple linear relation between the metal vapor pressure
$p_{\mathrm{M}}$ and the oxygen pressure $p_{\mathrm{O}_2}$:
\begin{equation}\label{eq-oxygen-balance-single}
p_{\mathrm{O}_2} = \frac{e}{c}\,p_{\mathrm{M}}
\end{equation}
Using Eq.~(\ref{eq-oxygen-balance-single}) one can now eliminate
$p_{\mathrm{O}_2}$ in Eq.~(\ref{eq-pvap-in-kequil}) in favor of
$p_{\mathrm{M}}$ and, with the known value of $K_{\mathrm{evap}}$
\modif{(Eq.~\ref{eq-equilconst-gibbs})}, solve for $p_{\mathrm{M}}$. 
\modif{This then leads to}
\begin{equation}\label{eq-pm-separate-congruent}
p_{\mathrm{M}} = \left(K_{\mathrm{evap}}\frac{a_{\mathrm{M}_c\mathrm{X}}}{{a_{\mathrm{Y}}}^d}
\left(\frac{c}{e}\right)^{e}\right)^{1/(e+c)}
\end{equation}
\modif{With Eq.~(\ref{eq-oxygen-balance-single})} this gives then also
$p_{\mathrm{O}_2}$. For the more minor metal oxide vapor species one can
solve Eq.~(\ref{eq-pvap-in-kequil}) now with the known value of
$p_{\mathrm{O}_2}$. If no single vapor species dominates, then
Eq.~(\ref{eq-oxygen-balance-single}) becomes a summation over all
contributing metals. An iteration procedure is then necessary, that
eventually converges in the value of $p_{\mathrm{O}_2}$.

Since we can evaluate $K_{\mathrm{evap}}$ for any given temperature, the
above procedure gives an expression for $p_{\mathrm{M}}$ in terms of the
activities of the metal oxide M$_c$X and (if applicable) of the residual
subtance Y. Typically, though, we specify as model parameter their mass
fraction $f$, not their activity $a$. We need to translate $f$ into $a$
before we can make use of Eq.~(\ref{eq-pvap-in-kequil}). This can be done
using the MELTS code\footnote{\url{http://melts.ofm-research.org}} (Ghiorso
\& Sack 1995). We define the {\em activity coefficient} $\gamma$ as
\begin{equation}
a = \gamma\, f
\end{equation}
where $f$ is the mass fraction\footnote{Conventionally the activity
  coefficient is defined in terms of the mole fraction, but for our purpose
  this definition is more convenient.} of the primary metal oxide, and $a$ is
the activity of the metal oxide species in the melt. For sodium this would
be
\begin{equation}\label{eq-act-coef-na}
a_{\mathrm{Na}_2\mathrm{SiO}_3}=\gamma\,f_{\mathrm{Na}_2\mathrm{O}}
\end{equation}
This $\gamma$ can be computed with MELTS for a specific composition (and
thus a specific $f$). It is, however, not always guaranteed that $\gamma$ is
independent of $f$. The MELTS code gives the activity for the dominant
compound involving the metal $M$ in the melt (e.g.\ for sodium:
Na$_2$SiO$_3$). Using equilibrium liquid $\leftrightarrow$ liquid reactions
the activity of the other compounds of the system can be easily
calculated. For instance, using the equilibrium reaction
\begin{equation}
\mathrm{Na}_2\mathrm{SiO}_3 (l) \leftrightarrow 
\mathrm{Na}_2\mathrm{O} (l) + \mathrm{SiO}_2 (l)
\end{equation}
one gets
$a_{\mathrm{Na}_2\mathrm{O}}=K_{\mathrm{eq}}a_{\mathrm{Na}_2\mathrm{SiO}_3}
/a_{\mathrm{SiO}_2}$, with the equilibrium constant $K_{\mathrm{eq}}$
computed using Eq.~(\ref{eq-equilconst-gibbs}). The equilibrium vapor
pressure $p_{\mathrm{Na}}$ computed for evaporation of Na$_2$SiO$_3$ will
then be identical to that computed for the evaporation of Na$_2$O. We
thus need to only compute a single evaporation channel.

\section{Equilibrium vapor pressures for Na, K, Fe, SiO and Mg}
\label{sec-peq-mgfenak}
Using the above procedure we now compute the equilibrium vapor pressure for
sodium in the temperature range between 1400 K and 2200 K. For the melt
composition we take composition 3 of Yu et al.~2003 (Geochimica et
Cosmochimica Acta 67, 773). Using the MELTS code we compute the activities,
and from that the activity coefficients. We define the activity coefficients
as
\begin{equation}
\begin{split}
\gamma_{\mathrm{Na}_2\mathrm{O}} = a_{\mathrm{Na}_2\mathrm{SiO}_3}/f_{\mathrm{Na}_2\mathrm{O}},\qquad
\gamma_{\mathrm{K}_2\mathrm{O}} = a_{\mathrm{KAlSiO}_4}/f_{\mathrm{K}_2\mathrm{O}},\\
\gamma_{\mathrm{Fe}\mathrm{O}} = a_{\mathrm{Fe}_2\mathrm{SiO}_4}/f_{\mathrm{Fe}\mathrm{O}},\qquad
\gamma_{\mathrm{SiO}_2} = a_{\mathrm{SiO}_2}/f_{\mathrm{SiO}_2},\\
\gamma_{\mathrm{Mg}\mathrm{O}} = a_{\mathrm{Mg}_2\mathrm{SiO}_4}/f_{\mathrm{Mg}\mathrm{O}},\qquad
\gamma_{\mathrm{Al}_2\mathrm{O}_3} = a_{\mathrm{Al}_2\mathrm{O}_3}/f_{\mathrm{Al}_2\mathrm{O}_3}\\
\end{split}
\end{equation}
We find that for MgO, SiO$_2$, FeO and K$_2$O these coefficients are almost
constant with temperature. For Al$_2$O$_3$ and Na$_2$O there is a mild
variation with temperature, but we will, for convenience, also assume them
to be constant with temperature, and take the value at 1850 K as
representative. The mass fractions and the computed activities and activity
coefficients are listed in Table \ref{tab-comp-yu-3}.

\begin{table}
\begin{center}
\begin{tabular}{l|rrr}
Species & $f_0$ \hspace{1.5em} & $a_0$ \hspace{1.5em} & $\gamma$ \hspace{1.5em}\mbox{} \\
\hline
Na$_2$O, Na$_2$SiO$_3$   &  2.4213E-02 &  6.8057E-04 &  2.8108E-02  \\
K$_2$O, KAlSiO$_4$       &  2.9260E-03 &  4.7220E-03 &  1.6138E+00  \\
FeO, Fe$_2$SiO$_4$       &  1.4427E-01 &  1.1313E-01 &  7.8418E-01  \\
SiO$_2$, SiO$_2$         &  5.5589E-01 &  5.6297E-01 &  1.0127E+00  \\
MgO, Mg$_2$SiO$_4$       &  1.8866E-01 &  2.0890E-01 &  1.1073E+00  \\
Al$_2$O$_3$, Al$_2$O$_3$ &  4.8426E-02 &  6.1667E-03 &  1.2734E-01  \\
\end{tabular}
\end{center}
\caption{\label{tab-comp-yu-3}The composition $f_0$ and activities $a_0$ of
  the relevant constituents for composition 3 of Yu et al.~(2003), where
  the Yu et al.\ mass fractions were normalized to unity (which leads to a
  small correction compared to the original Yu values). The first
  two columns list the compound of which the mass fraction $f_0$ is specified
  and the compound form for which the activity $a_0$ has been computed, respectively. 
  The activities were computed with the MELTS code and are valid at 1850K.  The
  activity coefficient $\gamma$ is also computed at 1850 K but turns out to
  be fairly constant with temperature.}
\end{table}

\begin{table}
\begin{center}
\begin{tabular}{l|rrrr}
Species & $q_0$ \hspace{1.5em} & $q_1$\hspace{1.5em} & $q_2$\hspace{1.5em} & $q_3$\hspace{1.5em}\mbox{}\\
\hline
%
%
Na    &  -3.3312E+01 &  3.2376E-02 & -1.2206E-05 &  1.6938E-09 \\
K     &  -3.2543E+01 &  2.6092E-02 & -7.4375E-06 &  6.6993E-10 \\
Fe    &  -5.6017E+01 &  5.7326E-02 & -2.2493E-05 &  3.2120E-09 \\
SiO   &  -6.2295E+01 &  5.9586E-02 & -2.1152E-05 &  2.7561E-09 \\
Mg    &  -6.1782E+01 &  5.7863E-02 & -2.0728E-05 &  2.7294E-09 \\
O$_2$ &  -3.4994E+01 &  3.4414E-02 & -1.3494E-05 &  1.9719E-09 \\
\end{tabular}
\end{center}
\caption{\label{tab-eqpress-yu-3}The resulting equilibrium vapor pressures
  for the composition of table \ref{tab-comp-yu-3} (Yu et al.\ 2003 
  composition 3) for the elements of interest (where all element 
  pressures are coupled via
  the common oxygen pressure), expressed as a fitting function
  $\mathrm{log}^{10}(p^{\mathrm{eq}})=q_0+q_1T+q_2T^2+q_3T^3$ with the
  pressure in bar and temperature in Kelvin. The fit was done at
  temperatures 1400, 1700, 2000 and 2200 K. These fits correspond
  to the curves in Fig.~\ref{fig-peq-all-coupled}.}
\end{table}

The evaporation reactions for the elements Na, K, Fe, SiO and Mg are then:
\begin{eqnarray}
\mathrm{Na}_2\mathrm{SiO}_3(l) &\leftrightarrow&
2\mathrm{Na}(g) + \tfrac{1}{2}\mathrm{O}_2(g) + \mathrm{SiO}_2(l) \\
2\mathrm{KAl}\mathrm{SiO}_4(l) &\leftrightarrow&
2\mathrm{K}(g) + \tfrac{1}{2}\mathrm{O}_2(g) + 2\mathrm{SiO}_2(l) 
+ \mathrm{Al}_2\mathrm{O}_3(l)\\
\mathrm{Fe}_2\mathrm{SiO}_4(l) &\leftrightarrow&
2\mathrm{Fe}(g) + \mathrm{O}_2(g) + \mathrm{SiO}_2(l) \\
\mathrm{SiO}_2(l) &\leftrightarrow&
\mathrm{SiO}(g) + \tfrac{1}{2}\mathrm{O}_2(g)\\
\mathrm{Mg}_2\mathrm{SiO}_4(l) &\leftrightarrow&
2\mathrm{Mg}(g) + \mathrm{O}_2(g) + \mathrm{SiO}_2(l)
\end{eqnarray}
These reactions allow us to calculate the equilibrium vapor pressures of the
elements \modif{for any given temperature $T$. We first calculate Gibbs free
  energy $\Delta_fG$ for each of the reactants and products using the
  polynomial fit of Eq.~(\ref{eq-gibbs-fit}) using the coefficients from
  Table \ref{tab-themo-coef}. Then from Eq.~(\ref{eq-deltarg}) we can
  calculate the change in the Gibbs free energy $\Delta_rG(T)$ for the
  reaction (where for the KAlSiO$_{4}$ reaction the equations have to be
  adapted accordingly). This then gives us the equilibrium coefficient
  $K_{\mathrm{evap}}$ using Eq.~(\ref{eq-equilconst-gibbs}). The activities
  of the liquid phases for our initial composition are obtained from Table
  \ref{tab-comp-yu-3}. From Eq.~(\ref{eq-pvap-in-kequil}) we now obtain the
  value of the product of vapor pressures
  ${p_{\mathrm{M}}}^c{p_{\mathrm{O}_2}}^e$ (where we recall that $c$ and $e$
  are defined in Eq.~\ref{eq-metalox-evap-reaction-general}). For congruent
  evaporation for each metal species separately we can express
  $p_{\mathrm{O}_2}$ directly in terms of $p_{\mathrm{M}}$ using
  Eq.~(\ref{eq-oxygen-balance-single}). The solution of $p_{\mathrm{M}}$
  is then found from Eq.~(\ref{eq-pm-separate-congruent}).}

\begin{figure}
\begin{center}
\includegraphics[width=0.47\textwidth]{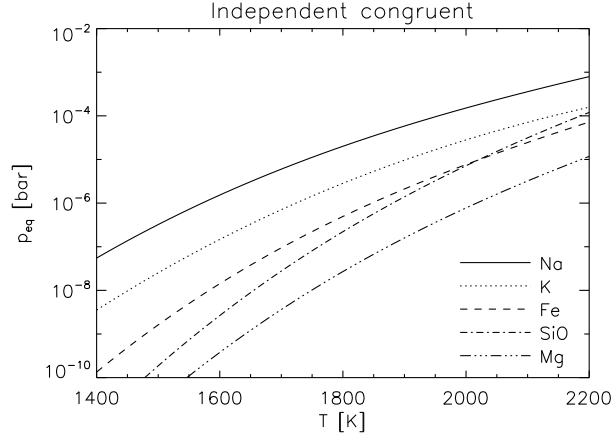}
\end{center}
\caption{\label{fig-peq-all-separate}The equilibrium vapor pressures
  calculated from the model for Na, K, Fe, SiO and Mg for composition 1 from Yu
  et al.~(2003). Each was calculated assuming a congruent oxygen pressure,
  assuming none of the other metal vapors to be present
  (i.e.~$p_{\mathrm{O}_2}=\tfrac{1}{4}p_{\mathrm{Na}}$,
  $p_{\mathrm{O}_2}=\tfrac{1}{4}p_{\mathrm{K}}$,
  $p_{\mathrm{O}_2}=\tfrac{1}{2}p_{\mathrm{Fe}}$,
  $p_{\mathrm{O}_2}=\tfrac{1}{2}p_{\mathrm{SiO}}$,
  $p_{\mathrm{O}_2}=\tfrac{1}{2}p_{\mathrm{Mg}}$ respectively).}
\end{figure}
\begin{figure}
\begin{center}
\includegraphics[width=0.47\textwidth]{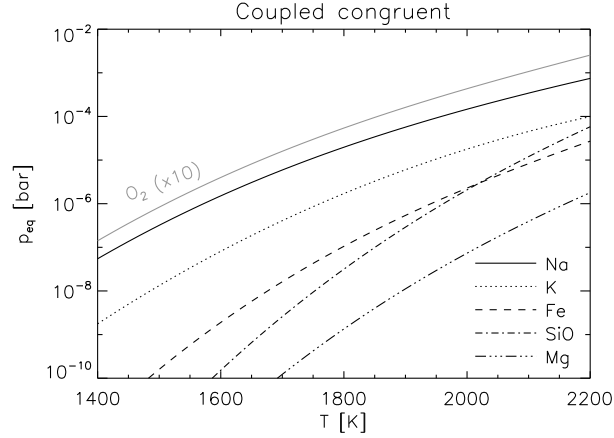}
\end{center}
\caption{\label{fig-peq-all-coupled}The equilibrium vapor pressures
  calculated from the model for Na, K, Fe, SiO and Mg for composition 1 from Yu
  et al.~(2003). They are all coupled by all contributing to the oxygen
  pressure according to Eq.~(\ref{eq-oxygen-adding-and-sharing}). Note that
  the oxygen pressure is shown at 10x its value, to avoid overcrowding of the
  figure.}
\end{figure}

\modif{A more self-consistent approach is to calculate} the vapor pressures
for the coupled system. In \modif{this case} case we iterate over the oxygen
pressure $p_{\mathrm{O}_2}$, where after each iteration we recompute it
according to
\begin{equation}\label{eq-oxygen-adding-and-sharing}
p_{\mathrm{O}_2} = \tfrac{1}{4}p_{\mathrm{Na}} + \tfrac{1}{4}p_{\mathrm{K}} +
\tfrac{1}{2}p_{\mathrm{Fe}} + \tfrac{1}{2}p_{\mathrm{SiO}}
+ \tfrac{1}{2}p_{\mathrm{Mg}}
\end{equation}
\modif{where we recall that we assume that there is no nebular gas inside
  the expanding cloud of molten droplets, and hence the oxygen pressure is
  entirely due to the evaporated material.}

In principle we should also consider the formation of a whole plethora of
diatomic molecules (such as MgO(g)) in the gas phase, which would add to the
vapor pressure (see e.g.~Fedkin et al.~2006). But these pressures are low
and we will ignore them.

For the above mentioned composition 3 of Yu et al.~(2003) the resulting
equilibrium pressures for congruent evaporation are shown in
Fig.~\ref{fig-peq-all-separate} for each species independently, and in
Fig.~\ref{fig-peq-all-coupled} for the coupled system. In Table
\ref{tab-eqpress-yu-3} these results are given in terms of the coefficients
of a third-order polynomial fit.

These results show that for this mixture (with a roughly 10x larger K
abundance than typically found in chondrules) the K pressure dominates. For
abundances of K more in line with typical chondrule values, the Na pressure
typically dominates. In the coupled figure one sees that the high oxygen
pressure set by K pushes the vapor pressures of Na, Fe and Mg down compared
to their independent values.

From the above analysis we see that the sodium equilibrium vapor pressure is
proportional to $\sqrt{a_{\mathrm{Na}_2\mathrm{SiO}_3}}$ for constant
$p_{\mathrm{O}_2}$ and proportional to
$a_{\mathrm{Na}_2\mathrm{SiO}_3}^{0.4}$ for congruent evaporation
(i.e.~$p_{\mathrm{O}_2}=0.25\,p_{\mathrm{Na}}$). Since the MELTS code finds
that the activity coefficient is approximately constant with the mass
fraction $f_{\mathrm{Na}_2\mathrm{O}}$ (at least for mass fractions not
higher than a few percent), it follows that
$p_{\mathrm{Na}}^{\mathrm{eq}}\propto \sqrt{f_{\mathrm{Na}_2\mathrm{O}}}$
for constant $p_{\mathrm{O}_2}$ and $p_{\mathrm{Na}}^{\mathrm{eq}}\propto
f_{\mathrm{Na}_2\mathrm{O}}^{0.4}$ for congruent evaporation. This is very
different from Raoult's law, and this has consequences for the
non-equilibrium behavior of evaporation, as we shall discuss in Section
\ref{sec-time-dep-evapcond-metal-oxides}.

Note that the square-root behavior of the sodium vapor pressure (for fixed
p$_{\mathrm{O}_2}$) is due to the fact that each unit of Na$_2$SiO$_3$ (or
each unit of Na$_2$O) in the liquid phase releases two units of Na
vapor. Alexander (2001) instead considers the basic unit in the liquid
to be NaO$_{1/2}$. This makes (at least for fixed $p_{\mathrm{O}_2}$) the
sodium vapor pressure again linear in the activity. This appears to be a
contradiction: two views of the same physics lead to different physical
predictions. This paradox is solved by realizing that the activity $a$ is
not necessarily linearly proportional to the mass fraction $f$.  For the
Na$_2$O versus NaO$_{1/2}$ paradox one can see this from the simple neutral
chemical ``reaction'' Na$_2$O $\leftrightarrow$ 2NaO$_{1/2}$, which has
$K_{\mathrm{eq}}=1=a_{\mathrm{NaO}_{1/2}}^2/a_{\mathrm{Na}_2\mathrm{O}}$
showing that $a_{\mathrm{NaO}_{1/2}}=\sqrt{a_{\mathrm{Na}_2\mathrm{O}}}$. So
if, as the MELTS code seems to indicate to some approximation,
$a_{\mathrm{Na}_2\mathrm{O}}\propto f_{\mathrm{Na}_2\mathrm{O}}$, then it
follows that $a_{\mathrm{Na}\mathrm{O}_{1/2}}\propto
\sqrt{f_{\mathrm{Na}_2\mathrm{O}}}$. Therefore both viewpoints lead to the
same behavior.

\section{Time-dependent evaporation/condensation of metal oxide melts}
\label{sec-time-dep-evapcond-metal-oxides}
\subsection{The Hertz-Knudsen equation and the definition of equilibrium vapor pressures under non-equilibrium conditions}
\label{sec-hkeq-and-definition-of-peq}
Let us now consider a drop of liquid that is time-dependently evaporating.
This process is governed by the Hertz-Knudsen equation:
\begin{equation}\label{eq-hertz-knudsen-eq-general}
\begin{split}
J &= J_{\mathrm{evap}} - J_{\mathrm{cond}} \\
&= v_{\mathrm{vap}}\left(\frac{\alpha_{\mathrm{evap}} p^{\mathrm{eq}}_{\mathrm{vap}}}{kT}-\alpha_{\mathrm{cond}}n_{\mathrm{vap}}\right)\\
&\equiv v_{\mathrm{vap}}\frac{\alpha_{\mathrm{evap}}p^{\mathrm{eq}}_{\mathrm{vap}}-\alpha_{\mathrm{cond}}p_{\mathrm{vap}}}{kT}
\end{split}
\end{equation}
where $J_{\mathrm{evap}}$ is the number of molecules per second per cm$^2$
leaving the surface of the drop, $J_{\mathrm{cond}}$ is the number of
molecules per second per cm$^2$ condensing onto the surface of the drop, $J$
is the net evaporation rate, $k$ is the Boltzmann constant, $T$ the
temperatur in Kelvin, $p^{\mathrm{eq}}_{\mathrm{vap}}$ is the equilibrium
vapor pressure (this time in units of dyne/cm$^2$), $n_{\mathrm{vap}}$ is
the number density of the vapor molecules in the gas phase (in units of
cm$^{-3}$), $p_{\mathrm{vap}} \equiv n_{\mathrm{vap}}kT$ is the vapor
pressure, $v_{\mathrm{vap}}\equiv \sqrt{kT/2\pi m}$ is the average projected
molecular velocity, $m$ is the molecular mass. The coefficients
$\alpha_{\mathrm{evap}}$ and $\alpha_{\mathrm{cond}}$ are the evaporation
and condensation coefficients, where $\alpha_{\mathrm{evap}}$ is the
probability that a molecule leaving the surface will remain in the vapor
phase and not get bounced back onto the surface by molecular collisions,
while $\alpha_{\mathrm{evap}}$ is the probability that a molecule that hits
the surface will stick to the surface and not bounce back into the gas.

To assure that thermodynamic equilibrium is guaranteed (i.e.\ that $J=0$
when $p_{\mathrm{vap}}=p_{\mathrm{vap}}^{\mathrm{eq}}$), we must have
$\alpha_{\mathrm{evap}}=\alpha_{\mathrm{cond}}$ under conditions near
equilibrium. Far from equilibrium, however, we could have
$\alpha_{\mathrm{evap}}\neq\alpha_{\mathrm{cond}}$. For the classical
evaporation process of molecules that do not dissociate upon evaporation
(e.g.~evaporation of water or methane) one can usually say that
$\alpha_{\mathrm{evap}}=\alpha_{\mathrm{cond}}$ and that
$p_{\mathrm{vap}}^{\mathrm{eq}}$ is independent of the conditions in the gas
above the liquid surface. That means then that the evaporation rate
$J_{\mathrm{evap}}$ is also independent on those conditions and saturation
is reached simply when the actual vapor pressure $p_{\mathrm{vap}}$ has
increased to the level of the equilibrium vapor pressure
$p_{\mathrm{vap}}^{\mathrm{eq}}(T)$.

For evaporation of metal oxide melts, where the evaporating metals are
accompanied by evaporating oxygen (i.e.\ dissociation upon evaporation), the
situation is more complex. It is still described by the Hertz-Knudsen
equation, this time for each of the vapor constituents separately (e.g.\ for
Na$_2$O we would write Eq.~\ref{eq-hertz-knudsen-eq-general} for Na vapor
and for O$_2$ separately, but clearly the equations are linked via
stoichiometry, because we must have
$J_{\mathrm{O}_2}=0.25\,J_{\mathrm{Na}}$), but now the equilibrium vapor
pressure, and therefore $J_{\mathrm{evap}}$, is no longer independent of the
conditions in the gas phase above the liquid surface. According to
Eq.~(\ref{eq-pvap-in-kequil}), if we increase the oxygen pressure
$p_{\mathrm{O}_2}$ in the gas above the liquid surface, then the equilibrium
vapor pressure for the metal vapor $p^{\mathrm{eq}}_{\mathrm{M}}$ decreases,
and therefore typically also its evaporation rate $J_{\mathrm{evap,M}}$.

Far away from equilibrium it becomes somewhat ambiguous how to define the
equilibrium vapor pressure. One way is to fix the oxygen pressure
$p_{\mathrm{O}_2}$ to the current value and compute the
$p^{\mathrm{eq}}_{\mathrm{M}}$ using Eq.~(\ref{eq-pvap-in-kequil}) for that
given value of $p_{\mathrm{O}_2}$. Another way is to assume congruent
evaporation in which the oxygen pressure is solved along with those of the
metal vapor pressures ($p^{\mathrm{eq}}_{\mathrm{O}_2}=\sum_i
(e_i/c_i)p^{\mathrm{eq}}_{\mathrm{M}_i}$), plus perhaps some background
oxygen pressure if present. Both definitions lead to different equilibrium
vapor pressures (compare Fig.~\ref{fig-peq-all-separate} with
Fig.~\ref{fig-peq-all-coupled}). Yet, the evaporation rate $J_{\mathrm{M}}$,
being a measureable quantity, should not be dependent on the choice of
definition. The different behavior of $\alpha_{\mathrm{evap,M}}$ and
$\alpha_{\mathrm{cond,M}}$ under these different definitions ensures that
the results are nevertheless identical.

Exactly what the values of $\alpha_{\mathrm{evap,M}}$ and
$\alpha_{\mathrm{cond,M}}$ are and how they change as a function of
$p_{\mathrm{M}}$, $p_{\mathrm{O}_2}$ (and other constituents of the gas
above the liquid surface) is not a simple question to answer. It
requires laboratory experiments under all the conditions of interest. The
complexity of all of this, in particular far away from equibrium conditions,
makes the numerical modeling of strongly non-equilibrium time-dependent
evaporation/condensation problems somewhat uncertain, unless the conditions
remain close to those measured in the laboratory.

Typically laboratory measurements are made of evaporation rates at low vapor
pressures, i.e.\ $p_{\mathrm{M}}\ll p_{\mathrm{M}}^{\mathrm{eq}}$. These
experiments are done either at given oxygen pressures or into near
vacuum. In some cases also other gases are present (e.g.~H$_2$ gas or air at
a given pressure). For some metals these additional gases can affect the
evaporation and condensation rates substantially. For instance, Yu et
al.~(2003) show that when H$_2$ gas is present, the evaporation rates of Na
are increased. From these measured values of
$J_{\mathrm{eqvap,M}}=J_{\mathrm{eqvap,M}}^{\mathrm{lab}}$, and for a given
choice of definition of the equilibrium vapor pressure, one can compute the
value of $\alpha_{\mathrm{evap,M}}$.

\subsection{Time-dependent vapor loss from molten metal oxides}
When a small molten lava droplet loses some of its constituents (say the
metal oxide M$_c$X) through the evaporation reaction 
\begin{equation}
\mathrm{M}_c\mathrm{X} (l) \leftrightarrow c\,\mathrm{M} (g) 
+ d\,\mathrm{Y} (l) + e\,\mathrm{O}_2 (g)
\end{equation}
into a {\em finite volume} $V$, the concentration of that constituent
declines and the vapor pressure of the corresponding metal gas increases. As
a result, the evaporation rate drops and the condensation rate rises. Let us
study the time-dependent loss of metal oxide M$_c$X (let us call this `the
volatile' from here on) from the droplet including both these
effects. Define the mass weighted abundance $f$ of this volatile inside the
lava droplet:
\begin{equation}\label{eq-mass-volatile-app}
m_{\mathrm{volatile}}=f\; m_{\mathrm{drop}}
\end{equation}
where $m_{\mathrm{volatile}}$ is the amount of gram of volatile that is
still remaining inside the droplet. We measure $f$ in terms of the
evaporating part of this metal oxide, e.g.\ for the evaporation of
Na$_2$SiO$_3$ we measure the mass fraction of Na$_2$O, because the rest is,
upon evaporation, left behind as SiO$_2$. We assume that the droplet is
completely liquid, so that the volatile is always perfectly mixed within the
droplet, and no concentration gradients form toward the surface. If we
ignore the difference in material density between the volatile and the rest
of the droplet, then the fraction of the surface that consists of volatile
particles is again $f$, meaning that $f$, together with the known 
saturation pressure for that volatile, indeed determines the evaporation
rate.

If we now have a finite volume $V$ around the droplet into which the vapor
can escape, then the vapor pressure $p_{\mathrm{M}}$ and the volatile mass
fraction $f$ are related via:
\begin{equation}\label{eq-relation-pm-ff0}
\frac{m_{\mathrm{vap}}Vp_{\mathrm{M}}}{kT}=(f_0-f)m_{\mathrm{drop}}
\end{equation}
where $f_0$ is the initial mass fraction \modif{of the volatile in the drop} before
evaporation and
\begin{equation}
m_{\mathrm{vap}} = m_{\mathrm{M}} + \frac{e}{c}m_{\mathrm{O}_2}
\end{equation}
is the mass of the metal atom plus the corresponding mass of the oxygen
molecules that join in the evaporation/condensation process. We have used
the ideal gas law $p_{\mathrm{M}}=n_{\mathrm{M}}kT$ \modif{(with $n_{\mathrm{M}}$
the number density of metal M vapor atoms in the gas phase)} combined with
the total mass of vapor (metal and oxygen) in the volume
$m_{\mathrm{vap}}Vn_{\mathrm{M}}$. Although \modif{the vapor pressure} $p_{M}$ and the
\modif{volatile mass fraction in the droplet} $f$ are related
through Eq.~(\ref{eq-relation-pm-ff0}), in the following we will still use
$p_M$ and $f$ separately for clarity.

The mass loss (or gain) rate of the droplet through evaporation (or
condensation) of the volatile metal oxide given by
\begin{equation}\label{eq-dmchondt-app}
\frac{d(m_{\mathrm{volatile}})}{dt} \equiv \frac{d(f m_{\mathrm{drop}})}{dt}
\equiv  m_{\mathrm{drop}}\frac{df}{dt} = -4\pi a_{\mathrm{drop}}^2J_{\mathrm{M}}m_{\mathrm{vap}}
\end{equation}
where $a_{\mathrm{drop}}$ is the radius of the droplet, $J_{\mathrm{M}}$ is
the evaporation rate of the metal M \modif{in units of number of vapor particles
per surface area}.

The evaporation rate of metal M can be expressed as a function of $f$ and
$p_{\mathrm{M}}$ using the Hertz-Knudsen equation
Eq.~(\ref{eq-hertz-knudsen-eq-general}), expressed here as
\begin{equation}\label{eq-hertz-knudsen-eq-complex}
  J_M(f,p_{\mathrm{M}},p_{\mathrm{O}_2},T)=\frac{\alpha_{\mathrm{evap,M}}\,
    p^{\mathrm{eq}}_{\mathrm{M}}(f,p_{\mathrm{O}_2},T)-
    \alpha_{\mathrm{cond,M}}\,p_{\mathrm{M}}}{\sqrt{2\pi m_{\mathrm{M}}kT}}
\end{equation}
where the evaporation and condensation coefficients 
$\alpha_{\mathrm{evap,M}}$ and $\alpha_{\mathrm{cond,M}}$ are
functions of $f$, $p_{\mathrm{M}}$, $p_{\mathrm{O}_2}$ and $T$.  In
principle all these functions can be non-linear functions, as described in
Sections \ref{sec-equil-pvap-metal-oxides} and
\ref{sec-hkeq-and-definition-of-peq}. The functional form of
the equilibrium vapor pressure 
$p^{\mathrm{eq}}_{\mathrm{M}}(f,p_{\mathrm{O}_2})$, once we fix its
definition (see Section \ref{sec-hkeq-and-definition-of-peq}), can be
computed using the math of Section \ref{sec-hkeq-and-definition-of-peq} and
a model of the activity $a_{\mathrm{M}_cX}$ as a function of the mass
fraction $f$ (e.g.\ the MELTS model). Typically this is a non-linear
behavior.

The functional form of the
$\alpha_{\mathrm{evap,M}}(f,p_{\mathrm{M}},p_{\mathrm{O}_2})$ and
$\alpha_{\mathrm{cond,M}}(f,p_{\mathrm{M}},p_{\mathrm{O}_2})$ coefficients
is not well known, due to limited experimental data, and generally requires
a simplified model or formula that is calibrated against the experimental
data that is available. There are numerous papers that describe the results
of such efforts (e.g.\ Alexander 2001, 2002; Fedkin et al.~2006, 2012, 2013
and many more). Typically a constant value is inferred by comparing the
measured evaporation rate $J$ at low pressures with
$\alpha_{\mathrm{evap,M}}p_{\mathrm{M}}^{\mathrm{eq}}$, and then it is
typically assumed that $\alpha_{\mathrm{cond,M}}=\alpha_{\mathrm{evap,M}}$.
One can use these values for modelling, as long as one remains close to the
conditions of the experiments against which these values were inferred. As
shown above, however, for conditions far from these, things become
uncertain.

Fedkin et al.~(2006) have done a careful study of several of these
experiments, and determined the $\alpha$ coefficients for Na, K, Mg, Fe and
SiO. Rather than doing our own fits to the experiments we use the values
reported in that paper. 

\section{Thermodynamic data}
\label{app-janafberman}
For most solid compounds we use the thermodynamic data from Ghiorso \& Sack
(1995), based on the Berman
model\footnote{\url{http://ctserver.ofm-research.org/ThermoDataSets/Berman.php}}
(Berman 1988). We also use the NIST JANAF
tables\footnote{\url{http://kinetics.nist.gov/janaf/}} (Chase 1998) for
several other compounds not in the Ghiorso \& Sack or Berman models, as
well as for the gas-phase.

Using thermodynamic data from different databases together in the same model
requires special care, because the definitions of the zero-enthalpy are
often different. For convenience of the reader we review here how the
databases are combined.

It is the international convention to define the formation enthalpy
$\Delta_f H^0(298)$ (in units of JK$^{-1}$mol$^{-1}$) of a certain compound
from its elements at the reference temperature $T_r=298.15$ K and at the
reference pressure of 1 bar. A compound is defined as a substance made of
multiple elements, such as SiO$_2$. The elements are defined as the pure
substances made of the elements or of the molecules defined as the reference
molecule. For the compound SiO$_2$ these are Si and O$_2$ (by convention the
reference oxygen is molecular oxygen, while for the metals the reference is
usually the pure metal: Fe, Mg, Na, Si etc). The formation enthalpy of the
elements in their reference state are therefore by definition $\Delta_f
H^0(298)=0$.

In the NIST-JANAF database tables the 6th column gives the
temperature-dependent formation enthalpy $\Delta_f H^0(T)$. This is defined
as the formation enthalpy at temperature $T$ from the most stable form of
the elements {\em at the same temperature}. This means that for a compound
this function is given by
\begin{equation}\label{eq-def-dfht}
\begin{split}
\Delta_f^{\mathrm{jnf}} H^0(T) = \Delta_f H^0(298) 
&+\left[H^0(T)-H^0(298)\right]_{\mathrm{compound}}\\
&-\sum_{i}\left[H^0(T)-H^0(298)\right]_{\mathrm{elem},i}
\end{split}
\end{equation}
(Chase 1998), ``jnf'' stands for NIST-JANAF, and $H^0(T)-H^0(298)$ is listed
in the 5th column of the tables. The formation Gibbs function using the same
definition is then
\begin{equation}
\begin{split}
\Delta_f^{\mathrm{jnf}} G^0(T) = \Delta_f^{\mathrm{jnf}} H^0(T) 
&-\left[TS^0(T)\right]_{\mathrm{compound}}\\
&+\sum_{i}\left[TS^0(T)\right]_{\mathrm{elem},i}
\end{split}
\end{equation}
(Chase 1998). One can verify that this is equal to column 7 of the NIST-JANAF
database. For the elements in their reference state the $\Delta_f H^0(T)$
and $\Delta_f G^0(T)$ are, by definition, zero at {\em all} temperatures.

In the Berman (1988) paper, as well as in the Holland \& Powell (1998) paper
and the Ghiorso \& Sack (1995) paper, the $\Delta_f H^0(298)$, the entropy
$S^0(298)$ and volume for 1 mole $V^0(298)$ are given at the reference
temperature. To compute $H^0(T)-H^0(298)$ one must integrate
\begin{equation}\label{eq-hhr-in-intcp}
H^0(T)-H^0(298) = \int_{298.15}^TC_P(T')dT'
\end{equation}
where $C_P(T)$ is the specific heat. An analytically integrable (polynomial)
fitting formula for $C_p(T)$ is given in Berman (1988) and a different one
in Holland \& Powell (1998). The coefficients of these fitting formulae are
listed in those papers. Berman (1988), Ghiorso \& Sack (1995) and Holland \&
Powell (1998) define the formation enthalpy $\Delta_f^{\mathrm{ber}} H^0(T)$
of a compound at temperature $T$ from its elements at the reference
temperature $T_r=298.15$:
\begin{equation}
\Delta_f^{\mathrm{ber}} H^0(T) = \Delta_f H^0(298) 
+\left[H^0(T)-H^0(298)\right]
\end{equation}
Let us, from now on, refer to the `Berman definition' whenever we mean
`Berman, Ghiorso \& Sack and Holland \& Powell definition'. The NIST-JANAF
and Berman definitions of the formation enthalpy are thus related as
follows:
\begin{equation}
\Delta_f^{\mathrm{jnf}} H^0(T) =\Delta_f^{\mathrm{ber}} H^0(T) 
-\sum_{i}\left[H^0(T)-H^0(298)\right]_{\mathrm{elem},i}
\end{equation}
In the Berman definition the formation Gibbs energy $\Delta_f^{\mathrm{ber}}
G^0(T)$ is also defined with respect to the elements at 298.15 K:
\begin{equation}
\begin{split}
\Delta_f^{\mathrm{ber}} G^0(T) &= \Delta_f^{\mathrm{ber}} H^0(T) -TS^0(T)\\
&=\Delta_f G^0(298) + \left[H^0(T)-H^0(298)\right]
-T\left[S^0(T)-S^0(298)\right]
\end{split}
\end{equation}
where $S^0(T)$ is found from $S^0(298)$ through
\begin{equation}\label{eq-s-in-intcpt}
S^0(T)-S^0(298) = \int_{298.15}^T\frac{C_P(T')}{T'}dT'
\end{equation}
which is also analytically integrable using the above mentioned fitting
polynomials. 
Note that in Berman's definition, neither $\Delta_f^{\mathrm{ber}} H^0(T)$
nor $\Delta_f^{\mathrm{ber}}G^0(T)$ are zero for temperatures $T\neq
298.15$K.  The NIST-JANAF and Berman definition of the formation Gibbs
function are related as follows:
\begin{equation}
\Delta_f^{\mathrm{jnf}} G^0(T) =\Delta_f^{\mathrm{ber}} G^0(T) 
-\sum_{i}\left[H^0(T)-H^0(298)\right]_{\mathrm{elem},i}
+\sum_{i}\left[TS^0(T)\right]_{\mathrm{elem},i}
\end{equation}
Of course the actual numbers are slightly different due to different sources
of the data and different approximations. 

For the computation of the equilibrium coefficient of a chemical reaction
(in our case evaporation) one can either use $\Delta_f^{\mathrm{jnf}}
G^0(T)$ or $\Delta_f^{\mathrm{ber}} G^0(T)$, as long as all substances
involved use the same definition, because the sum over the elements drops
out of the equation. However, when using data from both databases one must
first convert all data to the same convention.

In this paper we adopt the NIST-JANAF definitions. For convenience we make a
third-order polynomial fit to the dimensionless form of the Gibbs formation
energy $\Delta_f^{\mathrm{jnf}} G^0(T)/RT$ for several elements and
compounds of interest:
\begin{equation}\label{eq-gibbs-fit}
\frac{\Delta_f^{\mathrm{jnf}} G^0(T)}{RT} = a + b\,T + c\,T^2 + d\,T^3
\end{equation}
where $T$ is in Kelvin. The coefficients $a$, $b$, $c$ and $d$ are obtained
by fitting Eq.~(\ref{eq-gibbs-fit}) through the values at 1400, 1700, 2000
and 2200 Kelvin. These fits are good enough for the purpose of this paper
within the temperature range 1400 - 2200 K, but are not to be used beyond
this range since polynomial fits are known to quickly diverge when used
beyond their fitting range. The results are listed in Table
\ref{tab-themo-coef}.

\begin{table*}
\begin{center}
{\small
\begin{tabular}{l|rrrrr}
Species & a\hspace{2.5em} & b\hspace{2.5em} & c\hspace{2.5em} & d\hspace{2.5em} & Ref\mbox{}\\
\hline
SiO (g)             & -1.0517E+01 &-2.3503E-02 & 1.7198E-05 &-3.3390E-09 &JANAF    \\
Mg (g)              &  0.0000E+00 & 0.0000E+00 & 0.0000E+00 & 0.0000E+00 &JANAF    \\
Fe (g)              &  9.6481E+01 &-9.7974E-02 & 3.7134E-05 &-5.1464E-09 &JANAF    \\
Na (g)              &  0.0000E+00 & 0.0000E+00 & 0.0000E+00 & 0.0000E+00 &JANAF    \\
K (g)               &  0.0000E+00 & 0.0000E+00 & 0.0000E+00 & 0.0000E+00 &JANAF    \\
O$_2$ (g)           &  0.0000E+00 & 0.0000E+00 & 0.0000E+00 & 0.0000E+00 &JANAF    \\
Al$_2$O$_3$ (s/l)   & -4.1984E+02 & 3.8554E-01 &-1.4237E-04 & 1.9508E-08 &B+GS\\
SiO$_2$ (s/l)       & -1.9367E+02 & 1.5332E-01 &-4.7042E-05 & 5.2773E-09 &B     \\
MgO (s/l)           & -1.7563E+02 & 1.6901E-01 &-6.2570E-05 & 8.5805E-09 &JANAF    \\
Mg$_2$SiO$_4$ (s/l) & -5.5777E+02 & 4.9903E-01 &-1.7357E-04 & 2.2387E-08 &B+GS\\
MgSiO$_3$ (s/l)     & -3.8799E+02 & 3.4433E-01 &-1.1902E-04 & 1.5120E-08 &JANAF    \\
FeO (s/l)           & -7.4390E+01 & 7.6634E-02 &-3.1269E-05 & 4.6587E-09 &JANAF    \\
Fe$_2$O$_3$ (s)     & -1.7353E+02 & 1.5434E-01 &-4.9512E-05 & 5.4840E-09 &B+GS\\
Fe$_3$O$_4$ (s)     & -2.5146E+02 & 2.3429E-01 &-8.4549E-05 & 1.1470E-08 &JANAF    \\
Fe$_2$SiO$_4$ (s/l) & -3.3765E+02 & 3.0061E-01 &-1.0743E-04 & 1.4317E-08 &B+GS\\
FeSiO$_3$ (s)       & -2.6207E+02 & 2.1903E-01 &-7.1972E-05 & 8.8483E-09 &B   \\
Na$_2$O (s/l)       & -1.1988E+02 & 1.2844E-01 &-4.8364E-05 & 6.6846E-09 &JANAF    \\
Na$_2$SiO$_3$ (s/l) & -3.8065E+02 & 3.4204E-01 &-1.1879E-04 & 1.5348E-08 &B+GS\\
K$_2$O (s/l)        & -1.1465E+02 & 1.3492E-01 &-5.4470E-05 & 8.0269E-09 &JANAF    \\
KAlSiO$_4$ (s/l)    & -4.9643E+02 & 4.2598E-01 &-1.4312E-04 & 1.7709E-08 &B+GS\\
\end{tabular}
}
\end{center}
\caption{\label{tab-themo-coef}The coefficients for the polynomial fit to
  the dimensionless Gibbs free energy $\Delta_f^{\mathrm{jnf}}G(T)/RT$
  (Eq.~\ref{eq-gibbs-fit}) in the temperature range between 1400 and 2200
  K. The reference states of the elements are not listed because for them
  $\Delta_f^{\mathrm{jnf}}G(T)/RT=0$ by definition.  The coefficients were
  fitted to the data from the NIST-JANAF database (those with ref JANAF), 
  or the Berman (1998) model (ref B) or the Berman model for the solid
  phase and the Ghiorso \& Sack (1995) model for the liquid phase (ref B+GS).
  Note that for the solid phase Ghiorso \& Sack (1995) also use the
  Berman model. All the data in this table 
  are in the same definition (the JANAF definition) 
  and are thus mutually consistent. However, no guarantee is given that 
  the polynomial fits we show here are accurate
  enough for other applications than the simplified model of this paper.}
\end{table*}


\vspace{3em}


\begin{thebibliography}{100}
\bibitem{alexander:2001}Alexander, C.M.O., 2001, Meteoritics \& Planetary Science
36, 255
\bibitem{alexander:2002} Alexander, C.M.O, 2002, Meteoritics and Planetary Science, 37, 245
\bibitem{alexander:2008} Alexander, C.M.O., Grossman, J.N., Ebel, D.S., Ciesla, F.J., 
  2008, Science, 320, 1617
\bibitem{asphaug:2011} Asphaug, E., Jutzi, M., Movshovitz, N., 2011, 
  Earth and Planetary Science Letters, 308, 369
\bibitem{berman:1988} Berman, R.G., 1988, J.\ of Petrology, Vol 29, 445
\bibitem{chase:1998} Chase, M.W., 1998, Journal of physical and chemical
  reference data, Monograph No. 9
\bibitem{ciesla:2002} Ciesla, F.J., Hood, L.L., 2002, Icarus, 158, 281
\bibitem{cuzzi:2006} Cuzzi, J., Alexander, C.M.O., 2006, Nature, 441, 483
\bibitem{desch:2002} Desch, S.J., Connolly, H.C., 2002, Meteoritics and Planetary Science,
  37, 183
\bibitem{dullemond:2014} Dullemond, C.P., Stammler, S.M. and Johansen, A. 2014,
  ApJ, 794, 91 (Paper I)
\bibitem{Eisenhour:1995} Eisenhour, D.D. \& Buseck, P.R. 1995, Icarus, 117, 197–211
\bibitem{Fedkin:2006} Fedkin, A.V., Grossman, L. and Ghiorso, M.S., 2006, 
Geochimica et Cosmochimica Acta 70, 206
\bibitem{Fedkin:2012} Fedkin, A.V., Grossman, L., Ciesla, F.J. \& Simon, S.B. 
  2012, Geochimica et Cosmochimica Acta, 87, 81
\bibitem{Fedkin:2013} Fedkin, A.V. and Grossman, L., 2013, Geochimica et
  Cosmochimica Acta 112, 226–250
\bibitem{ghiorso:1995} Ghiorso, M.S.\ \& Sack, R.O., 1995, Contributions to
  Mineralogy and Petrology, 119, 197
\bibitem{gibbard:1997} Gibbard, S. G., Levy, E. H. \& Morfill, G. E. 1997, Icarus, 130, 517
\bibitem{hashimoto:1983} Hashimoto, A., 1983, Geochemical Journal 17, 111
\bibitem{hevey:2006} Hevey, P.J., Sanders, I.S., 2006, Meteoritics and Planetary Science,
  41, 95
\bibitem{hewins:2012b} Hewins, R.H.~\& Zanda, B.\ 2012,
Meteoritics \& Planetary Science, 47(7), 1120–1138
\bibitem{hewins:2012} Hewins, R.H., Zanda, B.\ \& Bendersky, C., 2012,
  Geochimica et Cosmochimica Acta, 78, 1
\bibitem{hewins:2005} Hewins, R.H., Connolly, H.C., Lofgren, G.E.\ \& Libourel, G., 
2005, in Chondrites and the Protoplanetary Disk, 341, 286
\bibitem{hollandpowell:1998} Holland, T.J.B. \& Powell, R., 1998, J.\ metamorphic Geol., 16, 309
\bibitem{hood:1991} Hood, L.L., Horanyi, M., 1991, Icarus, 93, 259
\bibitem{horanyi:1995} Horanyi, M., Morfill, G., Goertz, C.K. \& Levy,
  E.H. 1995, Icarus, 114, 174–185
\bibitem{hubbard:2012} Hubbard, A., McNally, C. P. \& Mac Low, M.-M. 2012, ApJ, 761, 58
\bibitem{jacquet:2012} Jacquet, E., Gounelle, M. \& Fromang, S. 2012, Icarus, 220, 162
\bibitem{johansen:2007} Johansen, A. \& Youdin, A. 2007, ApJ, 662, 627–641
\bibitem{johnson:2015} Johnson, B.C., Minton, D.A., Melosh, H. J. \& Zuber,
  M.T. 2015, Nature, 517, 339–341
\bibitem{kieffer:1975} Kieffer, S., 1975, Science, 189, 333
\bibitem{kingpring:2010} King, A.R. \& Pringle, J.E. 2010, MNRAS, 404, 1903–1909
\bibitem{krot:2005} Krot, A. N., Amelin, Y., Cassen, P. \& Meibom, A. 2005, Nature, 436, 989–992
\bibitem{morris:2010} Morris, M. A. \& Desch, S. J. 2010, The Astrophysical Journal, 722, 1474
\bibitem{sanders:2012} Sanders, I.S., Scott, E.R.D., 2012, Meteoritics and Planetary Science,
  47, 2170
\bibitem{sanders:2005} Sanders, I.S., Taylor, G.J., 2005, Chondrites and the Protoplanetary Disk,
  341, 915
\bibitem{shu:2001} Shu, F. H., Shang, H., Gounelle, M., Glassgold, A. E., Lee, T. 2001, ApJ, 548, 1029
\bibitem{stammler:2014} Stammler, S.M. \& Dullemond, C.P. 2014, Icarus 242, 1
\bibitem{urey:1953} Urey, H.C., Craig, H. 1953, Geochimica et Cosmochimica Acta, 4, 36
\bibitem{yu:2003} Yu, Y., Hewins, R.H., Alexander, C.M.O'D., and Wang, J., 2003
Geochimica et Cosmochimica Acta, 67, 773
\bibitem{zook:1980} Zook, H.A., 1980, Meteoritics, 15, 390
\end{thebibliography}
\end{document}